\documentclass[twocolumn]{aastex631}

\newcommand{\G}{\mathcal{G}}

\newcommand{\appropto}{\mathrel{\vcenter{\offinterlineskip\halign{\hfil$##$\cr\propto\cr\noalign{\kern2pt}\sim\cr\noalign{\kern-2pt}}}}}

\newcommand{\degree}{\ensuremath{\,^{\circ}}}

\usepackage{url}
\usepackage{graphicx}
\usepackage{amsmath,amstext}
\usepackage[T1]{fontenc}
\usepackage{breqn}

\usepackage{amsmath}
\usepackage{float}
\usepackage{longtable}
\usepackage{rotating}
\usepackage{tablefootnote}
\graphicspath{{./}{Figures/}}

\shorttitle{Obliquities x Orbits}
\shortauthors{Rice et al.}

\begin{document}
\title{The Orbital Geometries and Stellar Obliquities of Exoplanet-Hosting Multi-Star Systems}

\author[0000-0002-7670-670X]{Malena Rice}
\affiliation{Department of Astronomy, Yale University, New Haven, CT 06511, USA}

\author[0000-0002-4836-1310]{Konstantin Gerbig}
\affiliation{Department of Astronomy, Yale University, New Haven, CT 06511, USA}

\author[0000-0001-7246-5438]{Andrew Vanderburg}
\affiliation{Department of Physics and Kavli Institute for Astrophysics and Space Research, Massachusetts Institute of Technology, Cambridge, MA 02139, USA}

\correspondingauthor{Malena Rice}
\email{malena.rice@yale.edu}

\begin{abstract}
The current orbital geometries of exoplanet systems offer a fossilized record of the systems' dynamical histories. A particularly rich set of dynamical mechanisms is available to exoplanets residing in multi-star systems, which may have their evolution shaped by the gravitational influence of bound stellar companions. In this work, we examine the joint distribution of stellar obliquities and orbital orientations for transiting exoplanets residing within astrometrically resolved binary and triple-star systems. We leverage existing constraints on stellar obliquities in exoplanet systems, together with astrometric measurements from \textit{Gaia} DR3, to uncover a set of fully-aligned, ``orderly'' exoplanet systems that exhibit evidence of both spin-orbit and orbit-orbit alignment. We also find evidence that the observed distribution of orbit-orbit orientations in our sample is more strongly peaked toward alignment than an isotropic distribution. Our results may be indicative of efficient viscous dissipation by nodally recessing protoplanetary disks, demonstrating a regime in which stellar companions produce and maintain order in planetary systems, rather than enhancing misalignments.
\end{abstract}

\vspace{-10mm}

\section{Introduction} 
\label{section:intro}

A significant subset of all known exoplanet systems include a host star with one or more bound stellar companions \citep[e.g.][]{matson2018stellar, fontanive2021census}. These $N$-body systems can span a tremendous range of relative spin-orbit and orbit-orbit configurations, offering rich insights into the processes of star and planet formation. The geometries of these complex systems may reflect and inform the initial conditions within systems' natal molecular clouds \citep{bate2000predicting, bate2010chaotic, spalding2014alignment}, as well as long-term secular interactions with stellar companions \citep[e.g.][]{naoz2012formation, batygin2012primordial, lai2014star, naoz2016eccentric}.

Exoplanet spin-orbit angles, which serve as a proxy for stellar obliquities in exoplanet-hosting systems, provide an important lens into systems' dynamical histories. Spin-orbit angles can be measured using a range of methods reviewed in \citet{albrecht2022stellar}, with the most widespread methods being the Rossiter-McLaughlin effect \citep{rossiter1924detection, mcLaughlin1924some}, Doppler tomography \citep[e.g.][]{collier2010line, gandolfi2012doppler}, and starspot crossing events \citep[e.g.][]{sanchis2011starspots, dai2017stellar}. Each of these methods offers an avenue to constrain the sky-projected spin-orbit angle, $\lambda$, of a transiting exoplanet. Several past and ongoing surveys have collected spin-orbit measurements over recent decades \citep[e.g.][]{queloz2000detection, albrecht2011banana, albrecht2012obliquities, rice2021soles}, leading to the discovery of many heavily spin-orbit misaligned exoplanets \citep{albrecht2021preponderance} and culminating in a collection of spin-orbit measurements currently tracked within the TEPcat catalogue \citep{southworth2011homogeneous}.


The orbital orientations of multi-star systems reveal architectural insights that are complementary to those of spin-orbit orientations. Stellar binary and triple orbital orientations can be constrained through astrometric parallaxes and proper motion measurements \citep[e.g.][]{pearce2020orbital, tokovinin2020eccentricity, tokovinin2022resolved}. Using the \textit{Gaia} DR3 \citep{gaia2016, brown2022gaiadr3} catalogue of high-precision stellar astrometry, it is now possible to derive sky-projected stellar orbital orientations with substantially improved precision.

In this paper, we examine the joint distributions of spin-orbit and orbit-orbit orientations in multi-star systems hosting at least one transiting exoplanet. The work presented here is motivated by the recent finding that, at a population level, moderately wide binary systems that host transiting exoplanets are more often edge-on than field binaries, indicating a tendency toward orbit-orbit alignment in these systems \citep{christian2022possible, dupuy2022orbital}. Here, we add a new, complementary facet to this picture: the role of spin-orbit alignment in binary systems. 

We first demonstrate the existence of a population of ``orderly'', fully aligned systems that exhibit evidence of both spin-orbit and orbit-orbit alignment. We also find that the full distribution of orbit-orbit inclinations is more strongly peaked than an isotropic distribution, with a joint overabundance of aligned systems and dearth of orthogonal relative orbital orientations. 

Primordial orbit-orbit alignment may naturally arise in multi-star systems that form through disk fragmentation -- a mechanism in which one or more stellar companions are born from a gravitationally unstable circumprimary disk that inherits the primary's angular momentum vector \citep{Addams1989diskinst, BonnellBate1994binaryform}. This mechanism may help to explain the orbit-orbit aligned configurations observed for some shorter-period binary and multi-star systems \citep[e.g.][]{zhang2023mcdonald, lester2023visual}. However, the separations at which disk fragmentation can occur are bounded by the extent of the circumprimary disk. Disk fragmentation is, accordingly, expected to operate at relatively small companion separations $a\lesssim200$ au \citep{Krumholz2007, Tobin2016}, whereas the population probed in this work is dominated by moderate- to wide-separation systems.


As a result, we instead suggest that these observations provide evidence for efficient viscous dissipation during the protoplanetary disk phase, pushing systems toward orbit-orbit alignment while star-disk gravitational coupling maintains spin-orbit alignment. Our findings offer an observational counterpart to pre-existing theoretical frameworks that examine the influence of viscous dissipation on binary system alignment \citep{bate2000observational, lubow2000tilting, foucart2014evolution, Zanazzi2018}. 

Previous work has examined the range of available mechanisms through which stellar companions can excite the observed set of spin-orbit misalignments in exoplanet systems \citep[e.g.][]{naoz2012formation, batygin2012primordial, vick2023high}. By contrast, we identify and examine a regime of observed exoplanet systems in which stellar companions are more likely to have enhanced \textit{order} by pushing systems toward alignment.

\section{Methods}

\subsection{Initial sample selection}
We began by downloading the catalogue of all exoplanet spin-orbit angle measurements from TEPCat \citep{southworth2011homogeneous} on 12/30/2022. This catalogue, which includes a comprehensive, up-to-date listing of all published spin-orbit measurements, included 307 spin-orbit measurements obtained for 186 planets. Where multiple measurements were available for a given planet, the following criteria were applied to select a single measurement for our sample:

\begin{itemize}
     \item More recent publications that incorporated both new and previous data were favored.
    \item To maximize the uniformity of our sample, measurements made using either the Rossiter-McLaughlin effect or Doppler tomography were generally favored over spot-crossing measurements.
    \item For measurements that leveraged the Rossiter-McLaughlin effect, observations with full transit coverage -- and particularly those with pre- and post-transit baseline observations -- were favored.
    \item Measurements with no clear structure remaining in the Rossiter-McLaughlin residuals were favored.
\end{itemize}

Kepler-408 b was excluded from our sample because it has only a measured inclination, with no $\lambda$ measurement available \citep{kamiaka2019misaligned}. We also removed WASP-2 b from our sample because of its contested stellar obliquity measurement \citep{triaud2010spin, albrecht2011two}. In total, this left 184 individual planets\footnote{WASP-30 b has a sufficiently high mass ($M=60.86\pm0.89 M_J$) such that it may be more appropriately classified as a brown dwarf, rather than a planet \citep{anderson2011wasp30}. Nevertheless, we retained the system within our sample at this stage for completeness. No stellar companions were found for WASP-30 in Section \ref{subsection:characterizing_stellar_multiplicity}, such that this system was not included in the bulk of our analysis.} each with a sky-projected spin-orbit angle measurement.

\subsection{Collating sample properties}
We then cross-matched these 184 planets with the NASA Exoplanet Archive's Planetary Systems Composite Parameters table \citep[PSCompPars;][]{PSCompPars} to extract host star properties and coordinates, as well as planetary properties other than $\lambda$. Six of the 184 planets were not found in the PSCompPars table: TOI-778 b, TOI-1937 b, WASP-30 b, WASP-109 b, WASP-111 b, and WASP-134 b. For these planets, system properties were instead drawn from the Extrasolar Planet Encyclopaedia.\footnote{\url{http://exoplanet.eu/}}

Next, we cross-matched these systems with sources within the \textit{Gaia} DR3 catalogue using the RA, Dec ($\alpha$, $\delta$) coordinates of each host star and a search radius of $5\arcsec$. From this search, we obtained a single match to each of $149$ sources, two matches to each of $32$ sources, and three matches to each of $3$ sources. For systems with a single match, we adopted that \textit{Gaia} DR3 source as the host star. For systems with two or three matches, we cross-verified the host star's magnitude with candidate sources' stellar magnitudes to match the appropriate source to the host star of interest.

\subsection{Characterizing stellar multiplicity}
\label{subsection:characterizing_stellar_multiplicity}
Once the planet-hosting sources were identified in \textit{Gaia} DR3, we carried out a series of tests to identify bound stellar companions. In this way, we developed a robustly vetted sample of binary and triple-star systems for further study. No systems were found with more than three identified stellar components.

First, we conducted a cone search with radius $10\arcmin$ around each host star to search for bound stellar companions. We followed the methods of \citet{el2021million} to vet candidate companion sources (see Section 5.1 of \citet{Rice2023Qatar6} for a detailed overview of our vetting process). From this initial vetting step, we identified 44 host stars within our sample that reside in a system with at least one astrometrically observed stellar companion, including 43 tentative binary star systems and one triple-star system.

We cross-validated our identified systems with the binary catalogue derived in \citet{el2021million} and found that all 43 of our identified binary systems were also recovered within their study. All 43 identified pairs also have a low chance alignment probability $R<0.02$ as derived in \citet{el2021million}, with most systems having $R<1\times10^{-4}$.

Our initial vetting process excluded some host stars within our full sample that had both (1) a spin-orbit constraint and (2) one or more previously identified stellar companions that did not pass our tests. Through a literature search and cross-match with the NASA Exoplanet Archive, we identified 22 candidate multi-star systems in our sample, including 19 candidate binary star systems and three candidate triple-star systems, that had no bound companions identified by our first vetting step. We also found that three of our initially identified binary candidates (K2-290, Kepler-13, and WASP-24) are instead likely triple-star systems \citep{hjorth2019k2, mugrauer2019search}.

To investigate whether any of the 22 candidate multi-star systems could be included within our final sample, we examined the systems' astrometric fits, projected separations, and stellar magnitudes. The Renormalized Unit Weight Error (RUWE) parameter reported by \textit{Gaia} DR3 is a metric used to quantify how well a single-star model fits to the astrometric observations of a given source, where RUWE $\sim1.0$ signifies a good fit. Five of the 19 candidate binary systems (KOI-1257, WASP-2, WASP-20, WASP-72, and WASP-76) have primary stars with RUWE $>1.4$, suggesting that their stellar companions may be blended with the primary source and unresolvable with \textit{Gaia} astrometry. For another six candidate binary systems (TrES-2, HAT-P-14, HAT-P-27, HAT-P-29, HAT-P-32, and WASP-3) and two triple-star systems (HAT-P-8 and WASP-12), all stellar companions are are relatively faint and located within $\sim0.5-4\arcsec$ of the primary, such that they are likely also unresolved by \textit{Gaia} \citep{wollert2015lucky, ngo2015friends, belokurov2020unresolved, dupuy2022orbital}. These systems with no astrometrically constrained secondary sources could not be reincorporated into our final sample.

Five of the remaining systems (55 Cnc, HAT-P-20, EPIC 246851721, Kepler-89, and WASP-180) were recovered by loosening our parallax constraint, such that systems with relatively wide-separation (projected angular separation $\theta_s>4.0\arcsec$) companion candidates were accepted into the sample with parallaxes consistent up to $6\sigma$ (compared with our initial $3\sigma$ requirement adopted from \citet{el2021million}). One candidate triple-star system (WASP-11)\footnote{Only one stellar companion to WASP-11 was recovered within \textit{Gaia} DR3. However, WASP-11 has RUWE=3.60, indicating that the more nearby of its two stellar companions is likely unresolved.} and two candidate binary star systems (WASP-1 and WASP-33) were recovered by loosening our proper motion constraint, such that systems with relative proper motions consistent up to $5\sigma$ were accepted into the sample (compared with our initial $2\sigma$ requirement). This loosened proper motion constraint also recovered an additional binary system candidate -- HAT-P-70 -- which has not been previously flagged as a binary system in past publications. 

While a visual binary stellar companion to WASP-31 was previously identified at a relatively wide separation of $35\arcsec$ \citep{anderson2011wasp31}, we recovered no companion candidates that fit our looser constraints. WASP-31 has RUWE=0.99 and was not identified by \citet{el2021million} as a component of a stellar binary system. We also rule out the possibility that the binary star could have moved directly behind the primary during the time span between its initial detection and its observation by \textit{Gaia}, since a $35\arcsec$ separation corresponds to a projected distance of $\sim13,600$ au for a parallax $\varpi=2.5757$ mas from \textit{Gaia} DR3. We conclude that the previously identified visual binary companion was likely a chance alignment, rather than a bound companion.

In total, therefore, we recovered three initial binary candidates that were moved into our triple-star sample, eight additional candidate binary systems, and one additional candidate triple-star system. As expected based on our adopted constraints, none of the newly incorporated binary systems were listed within the binary star catalog developed by \citet{el2021million}. 

The systems that were not accepted into the sample using our original constraints are included in our ``loose'' sample, while they are excluded from our ``strict'' sample. Our ``loose'' sample includes 48 binary star systems and 5 triple-star systems, while our ``strict'' sample includes 40 binary star systems and 4 triple-star systems. The remainder of this work focuses on the more conservative ``strict'' sample in which all systems have passed our full vetting analysis. For thoroughness, we also consider the ``loose'' sample as a comparison point in Section \ref{subsection:significance_alignment}. All systems in each sample include a transiting planet with a spin-orbit measurement orbiting the primary star, as well as astrometric constraints from \textit{Gaia} DR3 for both the primary, planet-hosting star and at least one stellar companion.


\subsection{Astrometric orbital constraints}
\label{subsection:astrometric_orbital_constraints}

\begin{figure}
    \centering
    \includegraphics[width=0.48\textwidth]{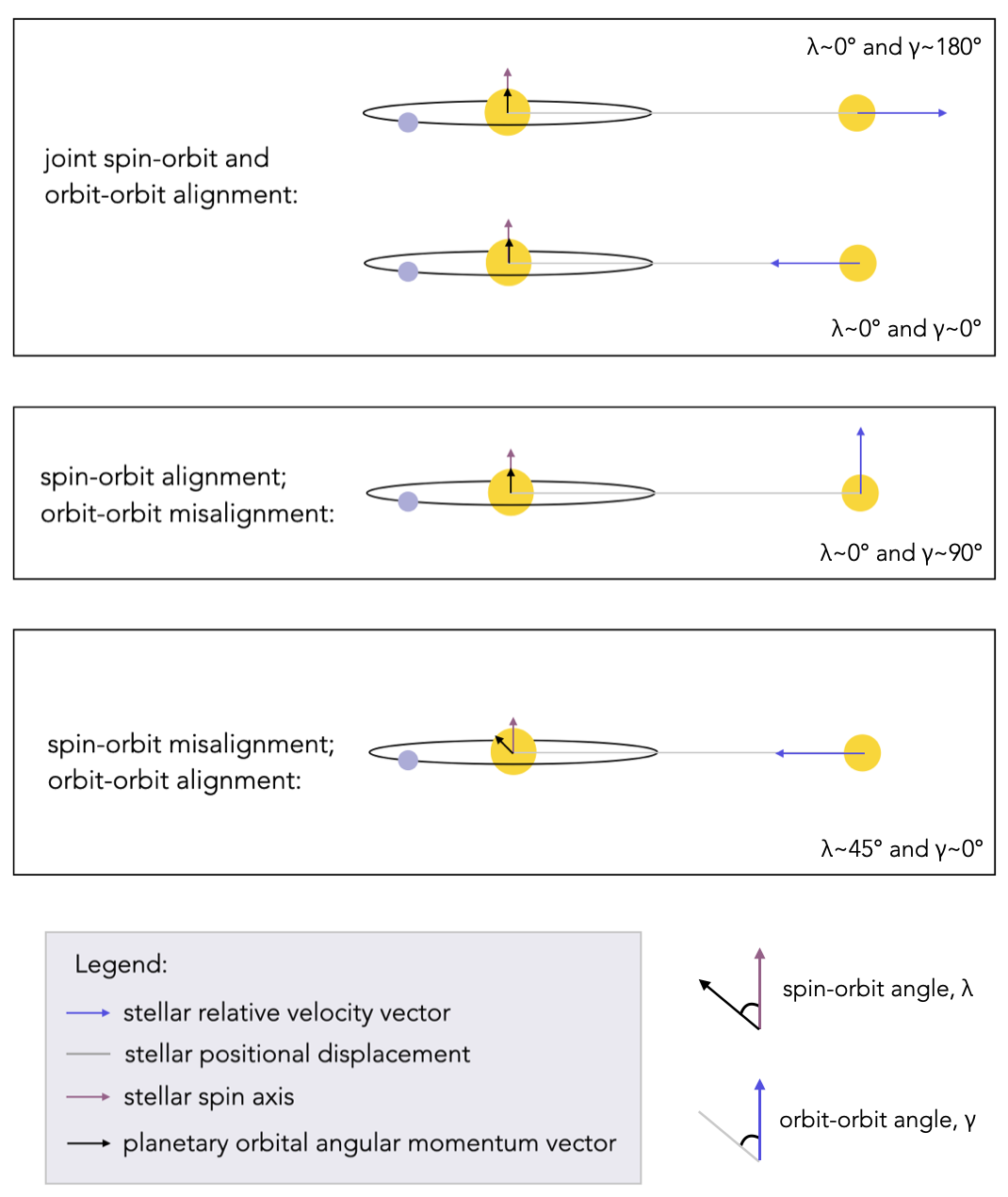}
    \caption{A selection of potential spin-orbit and orbit-orbit configurations for binary star systems with a circumprimary, transiting exoplanet. The transiting planet's orbit is shown in these example systems, while only the relative positional displacement and velocity vector are displayed for the stellar binary orbit.}
    \label{fig:binary_alignment_schematic}
\end{figure}

We next constrained the distribution of sky-projected stellar orbit configurations in our sample based on precise position and proper motion measurements from \textit{Gaia} DR3. Taken jointly with the existing set of exoplanet spin-orbit measurements, the relative stellar position and velocity vectors in multi-star systems can inform the range of full system geometries. Some potential configurations for exoplanet-hosting binary star systems are demonstrated in Figure \ref{fig:binary_alignment_schematic} for reference.

Using \textit{Gaia} DR3, we extracted the relative sky-projected position and velocity vectors $\vec{r}$ and $\vec{v}$ of the two stars within each binary system. The angle $\gamma$ between these two vectors is defined as

\begin{equation}
    \cos\gamma = \frac{\vec{r} \cdot \vec{v}}{|\vec{r}||\vec{v}|}
\end{equation}
where 

\begin{equation}
    \vec{r} \equiv [\Delta\alpha, \Delta\delta]
\end{equation}
and

\begin{equation}
    \vec{v} \equiv [\Delta\mu_{\alpha}^*, \Delta\mu_{\delta}].
\end{equation}

Here, $\Delta\alpha$ and $\Delta\delta$ are the positional differences between the primary and companion position in the RA and Dec directions, respectively, while $\Delta\mu_{\alpha}^*$ and $\Delta\mu_{\delta}$ are the corresponding proper motion differences in the RA and Dec directions. We note that \textit{Gaia} DR3 reports proper motions $\mu_{\alpha}^*$ that have already incorporated a $\cos\delta$ corrective factor in the RA direction, such that $\mu_{\alpha}^*\equiv\mu_{\alpha}\cos\delta$.\footnote{See the \texttt{pmra} definition in the \textit{Gaia} DR3 source documentation at \url{https://gea.esac.esa.int/archive/documentation/GDR3/Gaia_archive/chap_datamodel/sec_dm_main_source_catalogue/ssec_dm_gaia_source.html}.}

\begin{figure}
    \centering
    \includegraphics[width=0.48\textwidth]{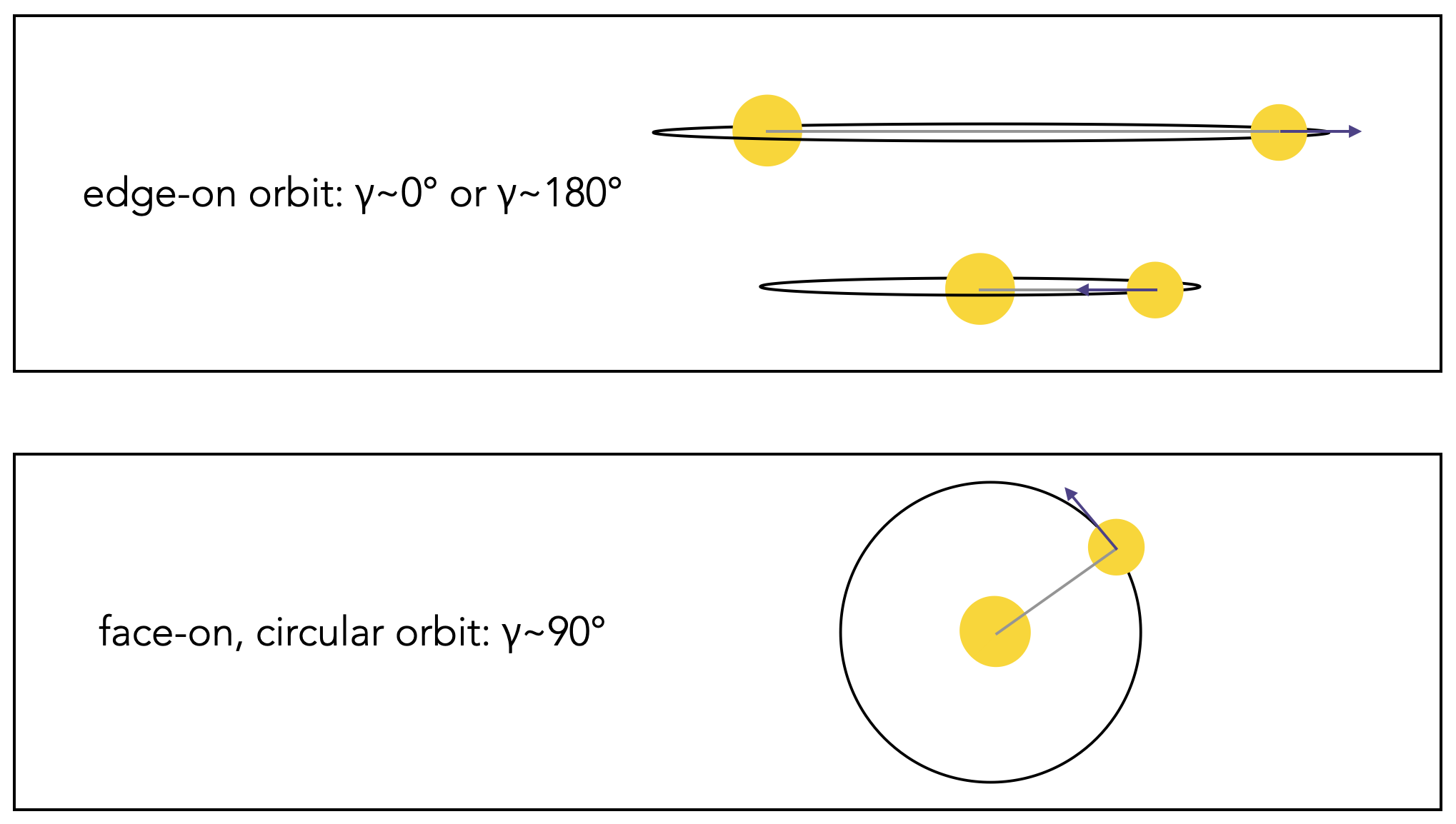}
    \caption{Two well-defined cases for $\gamma$: edge-on orbits ($\gamma\sim0\degree$ or $\gamma\sim180\degree$) and circular, face-on orbits ($\gamma\sim90\degree$). Here, the stellar binary orbit is shown for reference. Edge-on orbits always remain close to $\gamma\sim0\degree$ or $\gamma\sim180\degree$, whereas face-on orbits span a wide range of $\gamma$ values for nonzero eccentricities. The stellar relative velocity vector and positional displacement are colored here as in Figure \ref{fig:binary_alignment_schematic}.}
    \label{fig:gamma_visual}
\end{figure}

The angle $\gamma$ has been previously used to characterize the eccentricity distribution of wide stellar binary systems \citep{tokovinin2015eccentricity}. Two well-defined cases for $\gamma$ in binary star systems -- shown for edge-on orbits ($\gamma\sim0\degree$ or $\gamma\sim180\degree$) and for face-on, circular orbits ($\gamma\sim90\degree$) -- are demonstrated in Figure \ref{fig:gamma_visual} for reference. Systems with eccentric and/or moderately inclined orbits generally lie between these two extremes.

To derive our adopted $\gamma$ values and associated uncertainties, for each system we drew 30,000 random samples of each input variable -- $\alpha$, $\delta$, $\mu_{\alpha}^*$, and $\mu_{\delta}$ for both the primary and secondary star -- from \textit{Gaia} DR3. Parameter covariances were accounted for by multiplying the Cholesky decomposition of the correlation matrix describing these four variables with a normalized vector of uncorrelated Gaussian random variables. For most systems, this sampling produced a roughly Gaussian distribution of $\gamma$ values that was used to derive a new mean (adopted as $\gamma$) and standard deviation ($\sigma_{\gamma}$). 

The sampling for a few systems with $\gamma\sim0\degree$ or $\gamma\sim180\degree$ suffered from clear edge effects imposed by the definition of $\gamma$, which must exist within the range $0\degree\leq\gamma\leq180\degree$. For these systems, the sampled distribution emerged as a truncated Gaussian. In such cases, we found the peak of the Gaussian using a 100-bin histogram, and we mirrored the histogram across its central point before deriving $\gamma$ and $\sigma_{\gamma}$. The new, mirrored histogram allowed for values $\gamma<0\degree$ and $\gamma>180\degree$ for the purpose of deriving appropriate uncertainties. We verified that this mirroring method, when applied to non-truncated $\gamma$ distributions, well reproduced the mean and uncertainty values adopted from our original sampling. 

In four of our five identified triple-star systems, only one stellar companion was recovered through our \textit{Gaia} DR3 companion vetting search. This is likely because two of the stellar components have a small enough sky-projected separation such that the stars are blended within the \textit{Gaia} DR3 catalogue. Nevertheless, the configuration of two stellar components still offers useful geometric insights. Therefore, we calculated $\gamma$ for these triple-star systems in the same manner that we did for the binary star systems. For V1298 Tau, the triple-star system in our sample that has two resolved stellar components, we calculated $\gamma$ between the primary and each of the two stellar companions.




\subsection{3D orbit fitting for binary systems}
\label{subsection:orbit_fitting}
We also carried out 3D orbit fits to derive orbital inclinations for each binary star system using constraints drawn from \textit{Gaia} DR3. Deriving inclinations adds an extra layer of complexity relative to directly examining $\gamma$, which is a simple combination of direct observables from \textit{Gaia}. However, orbit forward modeling offers the advantage of explicitly accounting for the range of potential system eccentricities. 

To determine binary system inclinations, we first derived posterior distributions for the masses of all stellar companions identified within our search using the \texttt{isoclassify} Python package \citep{huber2017isoclassify, huber2017asteroseismology, berger2020gaia} in ``grid'' mode.  An all-sky dust model drawn from the \texttt{mwdust} Python package \citep{bovy2016galactic} was applied to account for extinction in all of our \texttt{isoclassify} models. For completeness, we included the triple-star system companions within our list of companions for which we derived masses; however, these were excluded from our inclination study, which was explicitly designed to fit binary system orbits.

All inputs to the \texttt{isoclassify} code were drawn from \textit{Gaia} DR3. These included stellar parallaxes; $G$, $Bp$, and $Rp$ magnitudes; and $T_{\rm eff}$, $\log g$, and [Fe/H] where available, as well as associated uncertainties. The \textit{Gaia} G band, which was available for all sources, was uniformly adopted as the model's reference absolute magnitude. An uncertainty floor of $5\%$ was adopted for $T_{\rm eff}$, $\log g$, and [Fe/H] -- a conservative bound loosely based upon the systematic stellar parameter uncertainty floors described in \citet{tayar2022guide}. A minimum photometric uncertainty of 1 mag was also set for each band of observations to ensure that the model grid converged on solutions for all systems. The results of this stellar companion mass derivation are provided in Table \ref{tab:companion_masses}.
\begin{deluxetable}{lll}
\tablecaption{Candidate stellar companion properties, with masses derived within this work. \label{tab:companion_masses}}
\tabletypesize{\scriptsize}
\tablehead{
\colhead{Companion name} & \colhead{RA, Dec ($^{\circ}$)} & \colhead{Mass ($M_{\odot}$)}}
\tablewidth{300pt}
\startdata
Binaries: & & \\
\hspace{2mm}55 Cnc B & (133.167832, 28.315252) & $0.30^{+0.10}_{-0.08}$ \\
\hspace{2mm}CoRoT-2 B & (291.776504, 1.382666) & $0.42^{+0.05}_{-0.07}$ \\
\hspace{2mm}CoRoT-11 B & (280.686736, 5.938083) & $0.96^{+0.36}_{-0.17}$ \\
\hspace{2mm}DS Tuc B & (354.914601, -69.194596) & $0.62\pm0.04$ \\
\hspace{2mm}EPIC 246851721 B & (78.918645, 16.277509) & $0.56^{+0.10}_{-0.12}$ \\
\hspace{2mm}HAT-P-1 B & (344.441521, 38.674035) & $1.10^{+0.14}_{-0.12}$ \\
\hspace{2mm}HAT-P-3 B & (206.097634, 48.027322) & $0.22^{+0.06}_{-0.05}$ \\
\hspace{2mm}HAT-P-4 B & (229.999935, 36.205037) & $1.10^{+0.12}_{-0.11}$ \\
\hspace{2mm}HAT-P-7 B & (292.248787, 47.969545) & $0.38^{+0.12}_{-0.11}$ \\
\hspace{2mm}HAT-P-16 B & (9.566866, 42.467720) & $0.79\pm0.05$ \\
\hspace{2mm}HAT-P-18 B & (256.346282, 33.011584) & $0.21^{+0.08}_{-0.06}$ \\
\hspace{2mm}HAT-P-20 B & (111.915081, 24.337609) & $0.58\pm0.11$ \\
\hspace{2mm}HAT-P-22 B & (155.683035, 50.131058) & $0.59\pm0.11$ \\
\hspace{2mm}HAT-P-24 B & (108.825344, 14.261251) & $0.44\pm0.11$ \\
\hspace{2mm}HAT-P-30 B & (123.949918, 5.837933) & $0.58^{+0.10}_{-0.11}$ \\
\hspace{2mm}HAT-P-41 B & (297.322581, 4.671410) & $0.78\pm0.12$ \\
\hspace{2mm}HAT-P-70 B & (74.547537, 9.968233) & $0.27^{+0.09}_{-0.08}$ \\
\hspace{2mm}HD 189733 B & (300.179016, 22.708365) & $0.37\pm0.01$ \\
\hspace{2mm}HD 80606 B & (140.665938, 50.603920) & $0.99^{+0.10}_{-0.08}$ \\
\hspace{2mm}K2-29 B & (62.671159, 24.402645) & $0.74^{+0.05}_{-0.06}$ \\
\hspace{2mm}KELT-9 B & (307.863809, 39.936958) & $0.17^{+0.06}_{-0.04}$ \\
\hspace{2mm}KELT-19 B & (111.505335, 7.605828) & $0.43\pm0.11$ \\
\hspace{2mm}KELT-24 B & (161.909223, 71.655152) & $0.61^{+0.12}_{-0.11}$ \\
\hspace{2mm}Kepler-25 B & (286.635515, 39.488638) & $0.77^{+0.05}_{-0.04}$ \\
\hspace{2mm}Kepler-89 B & (297.335668, 41.890478) & $0.72^{+0.05}_{-0.06}$ \\
\hspace{2mm}MASCARA-4 B & (147.577967, -66.114780) & $0.66^{+0.11}_{-0.10}$ \\
\hspace{2mm}Qatar-6 B & (222.210231, 22.151290) & $0.31^{+0.11}_{-0.08}$ \\
\hspace{2mm}TOI-1937 B & (116.370639, -52.382569) & $0.54^{+0.11}_{-0.14}$ \\
\hspace{2mm}TrES-1 B & (286.036339, 36.632911) & $0.16^{+0.06}_{-0.04}$ \\
\hspace{2mm}TrES-4 B & (268.304338, 37.212170) & $0.66^{+0.11}_{-0.10}$ \\
\hspace{2mm}WASP-1 B & (5.167002, 31.991244) & $0.37^{+0.11}_{-0.10}$ \\
\hspace{2mm}WASP-8 B & (359.901133, -35.032571) & $0.54^{+0.10}_{-0.12}$ \\
\hspace{2mm}WASP-14 B & (218.276917, 21.897881) & $0.20^{+0.07}_{-0.05}$ \\
\hspace{2mm}WASP-18 B & (24.350742, -45.684744) & $0.15\pm0.03$ \\
\hspace{2mm}WASP-26 B & (4.605675, -15.270848) & $0.79\pm0.05$ \\
\hspace{2mm}WASP-33 B & (36.679047, 37.573440) & $0.28^{+0.10}_{-0.08}$ \\
\hspace{2mm}WASP-49 B & (91.089756, -16.966021) & $0.41^{+0.16}_{-0.11}$ \\
\hspace{2mm}WASP-78 B & (63.767405, -22.121908) & $0.70\pm0.04$ \\
\hspace{2mm}WASP-85 B & (175.908426, 6.563716) & $0.87^{+0.16}_{-0.11}$ \\
\hspace{2mm}WASP-87 B & (185.327006, -52.842570) & $0.84^{+0.08}_{-0.06}$ \\
\hspace{2mm}WASP-94 B & (313.788298, -34.135730) & $1.14^{+0.10}_{-0.11}$ \\
\hspace{2mm}WASP-100 B & (68.959535, -64.028133) & $0.45\pm0.11$ \\
\hspace{2mm}WASP-111 B & (328.769230, -22.612847) & $0.65^{+0.11}_{-0.10}$ \\
\hspace{2mm}WASP-127 B & (160.547657, -3.836883) & $0.96^{+0.09}_{-0.08}$ \\
\hspace{2mm}WASP-167 B & (196.020884, -35.534295) & $1.01^{+0.15}_{-0.09}$ \\
\hspace{2mm}WASP-180 B & (123.393133, -1.983806) & $0.77\pm0.05$ \\
\hspace{2mm}WASP-189 B & (225.689198, -3.030630) & $0.52^{+0.11}_{-0.14}$ \\
\hspace{2mm}XO-2 S & (117.030969, 50.216885) & $0.98^{+0.09}_{-0.08}$ \\
Triples: & & \\
\hspace{2mm}K2-290 C & (234.857890, -20.202025) & $0.17^{+0.06}_{-0.04}$ \\ 
\hspace{2mm}Kepler-13 C & (286.970924, 46.868299) & $1.68^{+0.53}_{-0.43}$ \\ 
\hspace{2mm}V1298 Tau B & (61.309824, 20.139244) & $0.98^{+0.12}_{-0.11}$ \\ 
\hspace{2mm}WASP-11 C & (47.367252, 30.677693) & $0.61^{+0.49}_{-0.36}$ \\
\hspace{2mm}WASP-24 C & (227.209554, 2.342041) & $0.36^{+0.06}_{-0.07}$ \\
\enddata
\end{deluxetable}

Leveraging our newly measured stellar masses, we then used the \texttt{lofti\_gaia} Python package \citep{pearce2020orbital}, together with the \textit{Gaia} DR3 catalogue, to uniformly fit the 3D orbits of all binary systems within our sample. In brief, \texttt{lofti\_gaia} draws astrometric parameters from a selected \textit{Gaia} catalogue to probabilistically determine the allowed orbital properties of binary star systems using the Orbits For The Impatient (OFTI) Bayesian rejection sampling method \citep{blunt2017orbits}.

The user inputs to \texttt{lofti\_gaia} included the known primary star masses and uncertainties,\footnote{Primary star masses and uncertainties were set as the default parameter values in the NASA Exoplanet Archive PSCompPars table on 2023-02-09 \citep[][]{PSCompPars}.} as well as our newly derived secondary star masses and uncertainties (Table \ref{tab:companion_masses}). Where mass uncertainties were asymmetric, we adopted the larger uncertainty for use in \texttt{lofti\_gaia}. Each model fit was run up to 1000 accepted orbits, and the posterior distribution of orbital parameters was then drawn from the set of accepted solutions. The median inclination and the minimum and maximum 68\% credible interval for each system are reported as the central value and uncertainties, respectively, in Table \ref{tab:system_properties}.

\section{Results}

\begin{figure}
    \centering
    \includegraphics[width=0.48\textwidth]{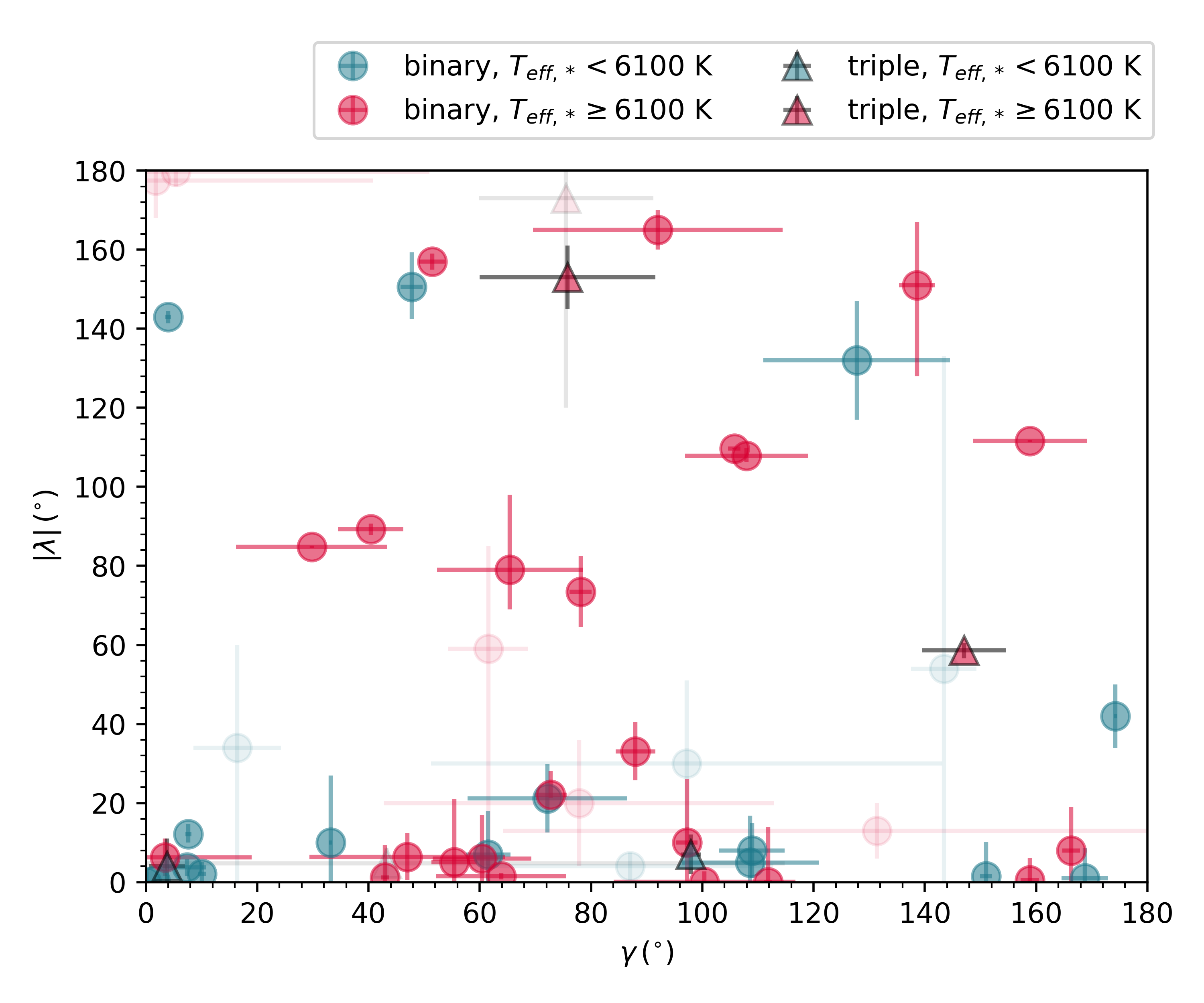}
    \caption{Sky-projected orbit-orbit angle ($\gamma$) vs. sky-projected spin-orbit angle ($\lambda$) for exoplanets in multi-star systems. For triple-star systems, $\gamma$ is measured between the primary and the nearest resolved stellar companion. Systems with $\sigma_{\gamma}>25\degree$ and $\sigma_{\lambda}>25\degree$ are shown at low opacity. Systems are colored by the host star's effective temperature, where the division between ``hot'' (red) and ``cool'' (blue) stars is set roughly at the location of the Kraft break.}
    \label{fig:gamma_v_lambda}
\end{figure}

\begin{figure}
    \centering
    \includegraphics[width=0.48\textwidth]{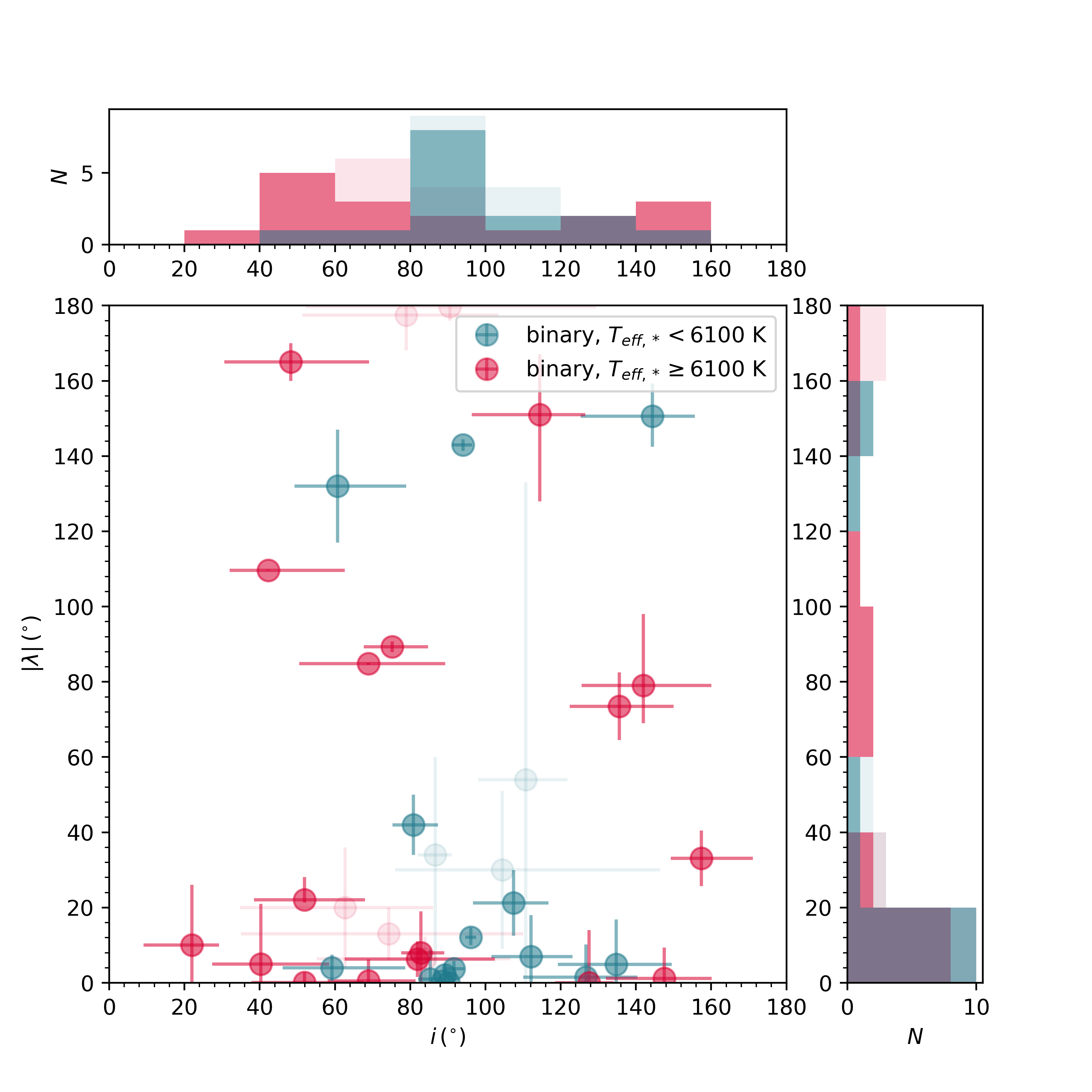}
    \caption{Stellar binary inclination ($i$) vs. sky-projected spin-orbit angle ($\lambda$) for exoplanets in binary star systems. Edge-on binary orbits correspond to $i=90\degree$, and all companion planets in the sample have near-edge-on orbits such that they transit the host star. Triple-star systems have been excluded from this figure. Systems with $\sigma_{i}>25\degree$ or $\sigma_{\lambda}>25\degree$ are shown at low opacity. Histograms of the 2D distributions of each parameter are provided in the top and right panels. Both the scatter plot and the histograms are colored by the host star's effective temperature, as in Figure \ref{fig:gamma_v_lambda}.}
    \label{fig:incl_v_lambda}
\end{figure}

The derived distribution of multi-star system orientations is shown in Figures \ref{fig:gamma_v_lambda} and \ref{fig:incl_v_lambda}, and the final set of systems and associated properties is provided in Table \ref{tab:system_properties}. Figure \ref{fig:gamma_v_lambda} shows the distribution of sky-projected spin-orbit angles $\lambda$ as a function of the sky-projected orbit-orbit angle $\gamma$ between the primary and secondary star, while Figure \ref{fig:incl_v_lambda} replaces $\gamma$ with the binary inclination $i$ as derived in Section \ref{subsection:orbit_fitting}. Triple-star systems are excluded from Figure \ref{fig:incl_v_lambda} because the orbit-fitting method used to extract inclinations is designed for binary star systems. Note that in Figure \ref{fig:gamma_v_lambda}, an edge-on orbit is signified by $\gamma\sim0\degree$ or $\gamma\sim180\degree$, whereas an edge-on orbit in Figure \ref{fig:incl_v_lambda} has $i\sim90\degree$. We discuss the implications of these key figures throughout the following sections.


\begin{deluxetable*}{llllllll}
\small
\tablecaption{Source IDs (SIDs) and adopted geometric properties for binary and triple-star systems examined within this work. For the triple-star systems, only the two closest-separation, \textit{Gaia}-resolved sources are listed. Sky-projected orbit-orbit angles $\gamma$ and inclinations $i$ were each derived within this work. Systems identified as ``fully aligned'' are denoted with an asterisk. Covariances between RA, Dec, and parallax were not considered when deriving uncertainties for each binary pair separation.}\label{tab:system_properties}
\tabletypesize{\scriptsize}
\tablehead{
\colhead{Host star} & \colhead{\textit{Gaia} DR3 SID, primary} & \colhead{\textit{Gaia} DR3 SID, cpn} & \colhead{sep (au)} & \colhead{$|\lambda| (\degree)$} & \colhead{Reference, $\lambda$} & \colhead{$\gamma (\degree)$} & \colhead{$i (\degree)$}}
\tablewidth{300pt}
\startdata
\multicolumn{7}{l}{} \\
Binaries: & & & & & & \\
\hspace{2mm}55 Cnc & 704967037090946688 & 704966762213039488 & $1065\pm1$ & $10^{+17}_{-20}$ & \citet{zhao2023measured} & $33.2\pm0.2$ & $68.1_{-8.1}^{+10.0}$\\
\hspace{2mm}*CoRoT-2 & 4287820848378092672 & 4287820852697823872 & $858\pm13$ & $1.0^{+7.7}_{-6.0}$ & \citet{czesla2012extended} & $168.8\pm4.4$ & $85.3_{-3.3}^{+4.2}$\\
\hspace{2mm}CoRoT-11 & 4285511294172309504 & 4285511294152417408 & $1749\pm384$ & $0.1\pm2.6$ & \citet{gandolfi2012doppler} & $100.3\pm15.4$ & $51.9_{-14.1}^{+15.6}$\\
\hspace{2mm}*DS Tuc A & 6387058411482257536 & 6387058411482257280 & $236\pm12$ & $12.1^{+2.6}_{-2.1}$ & \citet{benatti2021constraints} & $7.6\pm0.5$ & $96.1\pm1.5$\\
\hspace{2mm}EPIC 246851721 & 3393939030531019520 & 3393939026238329216 & $2922\pm732$ & $1.47^{+0.86}_{-0.87}$ & \citet{yu2018epic} & $63.9\pm11.7$ & $36.6_{-22.9}^{+15.1}$\\
\hspace{2mm}*HAT-P-1 & 1928431764627661440 & 1928431764627661312 & $1807\pm33$ & $3.7\pm2.1$ & \citet{johnson2008measurement} & $7.4\pm3.4$ & $91.6_{-2.5}^{+2.8}$\\
\hspace{2mm}HAT-P-3 & 1510191594552968832 & 1510191594552968960 & $1346\pm107$ & $21.2\pm8.7$ & \citet{mancini2018gaps} & $72.2\pm13.1$ & $107.5_{-10.8}^{+9.2}$\\
\hspace{2mm}HAT-P-4 & 1291120362349158016 & 1291119606434912384 & $29820\pm150$ & $4.9\pm11.9$ & \citet{winn2011orbital} & $108.4\pm10.7$ & $134.8_{-15.5}^{+14.7}$\\
\hspace{2mm}HAT-P-7 & 2129256395211984000 & 2129256395213382272 & $1362\pm2137$ & $177.5\pm9.4$ & \citet{winn2009hat} & $5.4\pm38.6$ & $79.0_{-27.7}^{+24.5}$\\
\hspace{2mm}HAT-P-16 & 381592313648387200 & 381592519806816896 & $5136\pm31$ & $10\pm16$ & \citet{moutou2011spin} & $97.2\pm1.9$ & $21.9_{-12.8}^{+7.3}$\\
\hspace{2mm}HAT-P-18 & 1334573817793362560 & 1334573817793501696 & $415\pm2178$ & $132\pm15$ & \citet{esposito2014gaps} & $127.8\pm17.7$ & $60.7_{-11.4}^{+18.3}$\\
\hspace{2mm}HAT-P-20 & 869913435026514688 & 869913430731341440 & $595\pm509$ & $8.0\pm6.9$ & \citet{esposito2017gaps} & $108.8\pm6.5$ & $26.0_{-12.4}^{+12.8}$\\
\hspace{2mm}*HAT-P-22 & 846946629987527168 & 846946625690867328 & $743\pm23$ & $2.1\pm3.0$ & \citet{mancini2018gaps} & $10.1\pm0.8$ & $89.2_{-0.6}^{+0.7}$\\ 
\hspace{2mm}HAT-P-24 & 3167323052618369408 & 3167323048322924672 & $2241\pm692$ & $20\pm16$ & \citet{albrecht2012obliquities} & $78.2\pm35.0$ & $62.7_{-27.9}^{+23.4}$\\
\hspace{2mm}HAT-P-30 & 3096441729861716224 & 3096441729861715968 & $808\pm55$ & $73.5\pm9.0$ & \citet{johnson2011hat} & $78.1\pm1.8$ & $135.6_{-13.2}^{+14.4}$\\
\hspace{2mm}HAT-P-41 & 4290415081653653632 & 4290415081653653376 & $1239\pm50$ & $22.1^{+6.0}_{-0.8}$ & \citet{johnson2017spin} & $72.7\pm2.9$ & $51.9_{-13.4}^{+16.0}$\\
\hspace{2mm}HAT-P-70 & 3291455819447952768 & 3291454921799634304 & $34993\pm5091$ & $107.9^{+2.0}_{-1.7}$ & \citet{bello2022mining} & $108.0\pm11.1$ & $115.1_{-30.1}^{+37.0}$\\
\hspace{2mm}*HD 189733 & 1827242816201846144 & 1827242816176111360 & $226\pm3$ & $0.4\pm0.2$ & \citet{cegla2016rossiter} & $1.7\pm0.1$ & $88.6\pm0.4$\\
\hspace{2mm}HD 80606 & 1019003226022657920 & 1019003329101872896 & $1354\pm7$ & $42\pm8$ & \citet{hebrard2010observation} & $174.3\pm0.3$ & $80.8_{-5.5}^{+6.5}$\\
\hspace{2mm}K2-29 & 150054788545735424 & 150054788545735296 & $775\pm69$ & $1.5\pm8.7$ & \citet{santerne2016k2} &  $151.0\pm1.0$ & $126.7_{-16.7}^{+13.6}$\\ 
\hspace{2mm}KELT-9 & 2064327278651198336 & 2064327205637457024 & $2679\pm342$ & $84.8\pm0.3$ & \citet{stephan2022nodal} & $30.0\pm12.9$ & $68.9_{-18.4}^{+20.4}$\\
\hspace{2mm}KELT-19 A & 3142847477107193344 & 3142846618113735552 & $11948\pm798$ & $179.7^{+3.8}_{-3.7}$ & \citet{siverd2017kelt} & $1.8\pm43.6$ & $90.6_{-38.4}^{+38.8}$\\
\hspace{2mm}KELT-24 & 1076970406751899008 & 1076970406752355584 & $194\pm253$ & $1.2^{+8.2}_{-7.4}$ & \citet{hjorth2019mascara} & $43.0\pm0.8$ & $147.5_{-15.5}^{+12.6}$\\
\hspace{2mm}Kepler-25 & 2100451630105041152 & 2100451630105040256 & $2045\pm56$ & $0.5\pm5.7$ & \citet{albrecht2013low} & $158.9\pm1.7$ & $68.9_{-10.8}^{+12.5}$\\
\hspace{2mm}Kepler-89 & 2076970047474270208 & 2076969841315840768 & $3620\pm258$ & $6^{+11}_{-13}$ & \citet{hirano2012planet}, & $60.4\pm8.6$ & $122.3_{-15.9}^{+11.3}$\\
\hspace{2mm}MASCARA-4 & 5245968236116294016 & 5245968236116983552 & $732\pm118$ & $109.66\pm0.14$ & \citet{zhang2022transmission} & $105.8\pm1.1$ & $42.3_{-10.2}^{+20.4}$\\
\hspace{2mm}*Qatar-6 & 1265513389372846848 & 1265513389372846720 & $478\pm102$ & $0.1\pm2.6$ & \citet{Rice2023Qatar6} & $0.2\pm1.6$ & $90.2_{-1.2}^{+1.1}$\\
\hspace{2mm}TOI-1937 & 5489726768531119616 & 5489726768531118848 & $1056\pm1834$ & $4.0\pm3.5$ & \citet{yee2023tess} & $86.6\pm25.8$ & $59.2_{-13.1}^{+19.5}$\\
\hspace{2mm}TrES-1 & 2098964849867337856 & 2098964884220974592 & $2465\pm904$ & $30\pm21$ & \citet{narita2007measurement} & $70.2\pm61.5$ & $104.5_{-28.5}^{+42.0}$\\
\hspace{2mm}*TrES-4 & 4609062308806929152 & 4609062381822880384 & $793\pm4008$ & $6.3\pm4.7$ & \citet{narita2010spin} & $4.5\pm14.8$ & $82.0_{-19.4}^{+20.5}$\\
\hspace{2mm}WASP-1 & 2862548428079638912 & 2862548428079638656 & $1539\pm1118$ & $59^{+26}_{-99}$ & \citet{albrecht2011two} &  $61.7\pm6.3$ & $22.5_{-14.7}^{+10.0}$\\
\hspace{2mm}WASP-8 & 2312679845530628096 & 2312679845529776128 & $406\pm52$ & $143.0^{+1.5}_{-1.6}$ & \citet{bourrier2017refined} & $4.0\pm0.5$ & $94.1_{-2.9}^{+2.4}$\\
\hspace{2mm}WASP-14 & 1242084170974175232 & 1242084166679297920 & $1894\pm89$ & $33.1\pm7.4$ & \citet{johnson2009third} & $88.0\pm3.4$ & $157.4_{-8.1}^{+13.7}$\\
\hspace{2mm}WASP-18 & 4955371367334610048 & 4931352153572401152 & $2959\pm804$ & $13\pm7$ & \citet{albrecht2012obliquities} & $115.2\pm56.3$ & $74.3_{-39.3}^{+35.7}$\\
\hspace{2mm}WASP-26 & 2416782701664155008 & 2416782705960292608 & $3890\pm34$ & $34^{+26}_{-36}$ & \citet{albrecht2012obliquities} & $16.0\pm8.6$ & $86.6_{-4.6}^{+4.4}$\\
\hspace{2mm}WASP-33 & 328636019723252096 & 328636264539126912 & $18254\pm2608$ & $111.64\pm0.28$ & \citet{watanabe2022nodal} & $159.0\pm10.2$ & $68.0_{-39.4}^{+35.6}$\\
\hspace{2mm}WASP-49 & 2991284369063612928 & 2991284162905572992 & $433\pm543$ & $54^{+79}_{-58}$ & \citet{wyttenbach2017hot} & $143.4\pm6.0$ & $110.7_{-12.6}^{+11.1}$\\
\hspace{2mm}WASP-78 & 5089851638095503616 & 5089851638095502592 & $30995\pm588$ & $6.4\pm5.9$ & \citet{brown2017rossiter} & $48.9\pm17.0$ & $84.0_{-28.8}^{+22.7}$\\
\hspace{2mm}WASP-85 A & 3909745223886018432 & 3909745223886018560 & $206\pm92$ & $0\pm14$ & \citet{mocnik2016starspots} & $111.8\pm0.6$ & $127.5_{-9.0}^{+6.6}$\\ 
\hspace{2mm}WASP-87 & 6077185317188247936 & 6077185317188247552 & $2482\pm76$ & $8\pm11$ & \citet{addison2016spin} & $166.3\pm2.0$ & $82.9_{-5.2}^{+6.2}$\\
\hspace{2mm}WASP-94 A & 6780546169634170496 & 6780546169633474944 & $3190\pm31$ & $151^{+16}_{-23}$ & \citet{neveu2014wasp} & $138.6\pm3.7$ & $114.5_{-18.1}^{+12.0}$\\
\hspace{2mm}WASP-100 & 4675352109659600000 & 4675352109658261120 & $1353\pm1076$ & $79^{+19}_{-10}$ & \citet{addison2018stellar} & $65.4\pm13.3$ & $142.0_{-16.4}^{+18.0}$\\
\hspace{2mm}WASP-111 & 6813902839862151936 & 6813902839862983296 & $1469\pm222$ & $5\pm16$ & \citet{anderson2014six} & $55.4\pm4.6$ & $40.3_{-13.0}^{+18.2}$\\
\hspace{2mm}WASP-127 & 3778075717162985600 & 3778075717162986240 & $6514\pm14$ & $150.6^{+8.7}_{-8.1}$ & \citet{cristo2022carm} & $47.8\pm2.1$ & $144.4_{-19.0}^{+11.3}$\\
\hspace{2mm}WASP-167 & 6154982877300947840 & 6154960229936153344 & $35440\pm319$ & $165.0\pm5.0$ & \citet{temple2017wasp} & $93.1\pm23.0$ & $48.2_{-17.6}^{+20.9}$\\
\hspace{2mm}WASP-180 A & 3070964117005132416 & 3070964117005860736 & $1265\pm15$ & $157.0\pm2.0$ & \citet{temple2019wasp} & $51.5\pm2.4$ & $120.8_{-12.0}^{+11.2}$\\
\hspace{2mm}WASP-189 & 6339097679918871168 & 6339097675623770496 & $959\pm15$ & $89.3\pm1.4$ & \citet{anderson2018wasp} & $40.4\pm5.7$ & $75.2_{-7.5}^{+9.6}$\\
\hspace{2mm}XO-2 N & 934346809278715776 & 934346740559239296 & $4677\pm15$ & $7\pm11$ & \citet{damasso2015gaps_xo2} & $61.5\pm4.0$ & $112.2_{-10.5}^{+11.0}$\\
Triples: & & & & & & \\
\hspace{2mm}K2-290 & 6253844468883543680 & 6253844464585162880 & $2735\pm272$ & $153\pm8$ & \citet{hjorth2021backward} & $75.2\pm16.2$ & - \\
\hspace{2mm}Kepler-13 & 2130632159136095104 & 2130632159130638464 & $552\pm2687$ & $58.6\pm2.0$ & \citet{masuda2015spin} & $147.2\pm7.6$ & - \\
\hspace{2mm}*V1298 Tau & 51886335968692480 & 51884824140205824 & $10515\pm26$ & $4^{+7}_{-10}$ & \citet{johnson2022aligned} & $4.2\pm3.2$ & - \\
\hspace{2mm}WASP-11 & 123376685084303360 & 123376719445431552 & $2076\pm62$ & $7\pm5$ & \citet{mancini2015gaps} & $97.9\pm1.7$ & - \\
\hspace{2mm}WASP-24 & 1153682508388170112 & 1153682508388170496 & $7117\pm178$ & $4.7\pm4.0$ & \citet{simpson2011spin} & $66.6\pm55.5$ & - \\
\enddata
\tablecomments{Only one of the two $\lambda$ measurements made in the K2-290 system is provided here; the second measurement, for the inner planet K2-290 b, is significantly less well-constrained at $\lambda=173^{+45}_{-53}\degree$ \citep{hjorth2021backward}.}
\end{deluxetable*}

\subsection{A population of fully aligned systems}
\label{subsection:aligned_systems}

A primary result of this work is the discovery of an overabundance of systems that are consistent with joint orbit-orbit and spin-orbit alignment. We identify seven binary star systems and one triple-star system that robustly fit into this group, shown in the bottom left corner of Figure \ref{fig:gamma_v_lambda} and the bottom center of Figure \ref{fig:incl_v_lambda}. All seven systems are part of our ``strict'' sample and have:

\begin{enumerate}
    \item a small sky-projected spin-orbit angle consistent with $|\lambda|\leq10\degree$ within $1\sigma$;
    \item a sky-projected orbit-orbit angle $\gamma$ within $10\degree$ of either $0\degree$ or $180\degree$ at the $1\sigma$ level, such that $90\degree - |90\degree-\gamma|\leq10\degree$ between the planet-hosting star and its \textit{Gaia}-resolved stellar companion;
    \item binary inclinations within $10\degree$ of an edge-on configuration, such that $|90\degree - i|\leq10\degree$ within $1\sigma$;
    \item uncertainties $\sigma_{\lambda}<25\degree$, $\sigma_{\gamma}<25\degree$, and, for binary star systems, $\sigma_{i}<25\degree$.
\end{enumerate}

One fully aligned binary system candidate within this population has been previously identified and characterized: Qatar-6 \citep{Rice2023Qatar6}. In this work, we have identified six additional stellar binary systems that each show evidence of both spin-orbit and orbit-orbit alignment -- CoRoT-2, DS Tuc, HAT-P-1, HAT-P-22, HD 189733, and TrES-4. We have also identified one triple-star system, V1298 Tau, that hosts a spin-orbit aligned exoplanet \citep{johnson2022aligned, gaidos2022zodiacal} and that exhibits a robust line-of-sight orbit-orbit alignment between the primary and secondary stellar components (the alignment does not extend to the tertiary stellar component). The orbit-orbit alignment in each of these seven systems has been identified for the first time in this work. The V1298 Tau primary star hosts two co-transiting exoplanet companions that are both consistent with spin-orbit alignment.

While this information was not explicitly taken into consideration in our selection criteria, we note that the 3D spin-orbit angles $\psi$ have also been reported for CoRoT-2 b ($\psi=8.9^{+6.7}_{-5.1}\degree$), DS Tuc b ($\psi=4.0^{+4.6}_{-1.6}\degree$), HAT-P-22 b ($\psi=6.7^{+29.4}_{-3.8}\degree$), and HD 189733 b ($\psi=2.3^{+13.5}_{-1.6}\degree$) in \citet{albrecht2021preponderance}; for V1298 Tau b ($\psi=8^{+4}_{-7}\degree$) in \citet{johnson2022aligned}; and for Qatar-6 b ($\psi=21.8^{+8.9}_{-18.4}\degree$) in \citet{Rice2023Qatar6}. All are consistent with alignment, with $\psi\leq10\degree$ within $1\sigma$.

Independent work based on long-term astrometric orbital monitoring recently found one other system, TrES-2, that also demonstrates full spin-orbit and orbit-orbit alignment \citep{dupuy2022orbital}. The TrES-2 system was removed from our sample at the vetting stage due to the faint magnitude and small projected separation of the companion star ($\rho_0=1.1\arcsec$ from \citet{dupuy2022orbital}, which corresponds to sky-projected separation $s=239$ au when combined with the host star's parallax from \textit{Gaia} DR3), such that it was not resolved in the \textit{Gaia} dataset. This independent discovery offers complementary evidence toward the observed overabundance of aligned systems.

\begin{figure}
    \centering
    \includegraphics[width=0.48\textwidth]{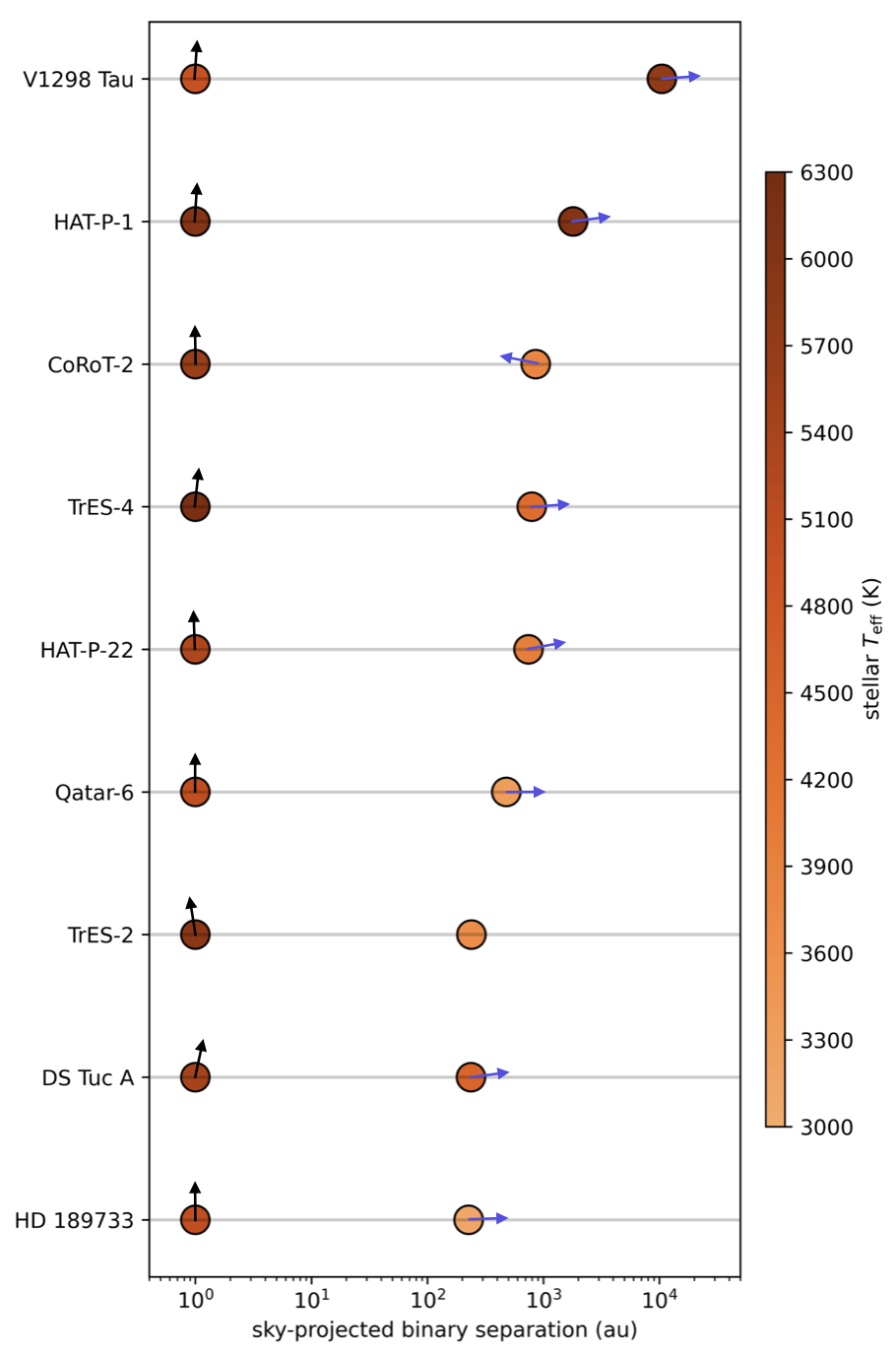}
    \caption{The nine identified ``fully aligned'' systems, ordered by the sky-projected separation between the exoplanet host star and its aligned companion star. The same color scheme from Figure \ref{fig:binary_alignment_schematic} was used here, with stellar relative velocity vectors shown in blue for all systems with $\gamma$ measured within this work. All primary, exoplanet-hosting stars are placed at the 1 au mark, with the sky-projected stellar spin axis (as approximated by $\lambda$) shown for each host star. Temperatures were drawn from the NASA Exoplanet Archive for exoplanet host stars, and they were drawn either from \textit{Gaia} DR3 or from previously published works \citep{ngo2015friends, mugrauer2019search} where available for the secondary stars. No previous measurement of $T_{\mathrm{eff}}$ was found for TrES-4 B, so the value was estimated from our isochrone fit as $T_{\mathrm{eff}}=4349^{+521}_{-386}$ K.}
    \label{fig:aligned_systems}
\end{figure}

This group of nine systems -- including our seven identified fully aligned binary system candidates, TrES-2, and V1298 Tau -- comprises a newly discovered population of systems consistent with full spin-orbit and orbit-orbit alignment between the primary and secondary stellar components. The configurations of these systems are shown in Figure \ref{fig:aligned_systems}. Eight of the nine identified systems include an exoplanet host star below the Kraft break ($T_{\rm eff}\sim6000-6250$ K), a rotational discontinuity above which stars have radiative envelopes \citep{kraft1967studies}. As a result, their spin-orbit alignment is consistent with either (1) a primordial coupling between the spin axis and the natal protoplanetary disk \citep{spalding2014alignment, spalding2015magnetic} or (2) tidal realignment through dissipation in the convective envelope of the cool host star \citep{winn2010hot, albrecht2012obliquities, rice2022origins}.

The fully aligned sample spans a range of sky-projected stellar separations. Most fully aligned systems have a sky-projected separation $s=200-900$ au between the primary and the secondary star, with two notable exceptions: HAT-P-1, with a sky-projected binary separation $s=1,806$ au, and V1298 Tau, the aligned triple-star system within our sample, with $s=10,515$ au between its two aligned components. 



\begin{figure}
    \centering
    \includegraphics[width = \linewidth]{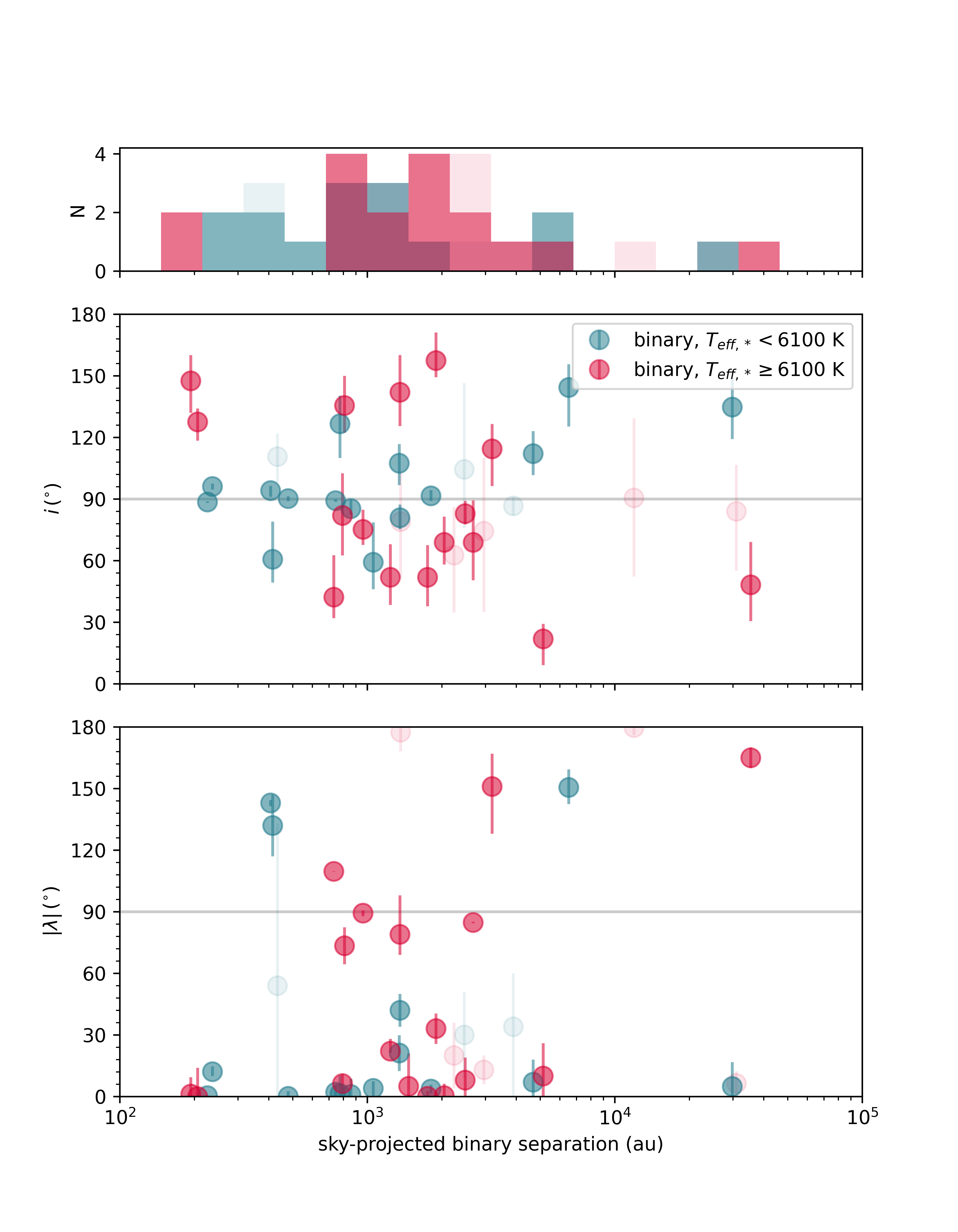}
    \caption{Binary inclination $i$ (top panel) and sky-projected spin-orbit angle $|\lambda|$ (bottom panel) as a function of sky-projected binary separation. The gray lines indicate $i = 90^\circ$ and $|\lambda| = 90^\circ$ for reference. As in Figure \ref{fig:incl_v_lambda}, systems with $\sigma_i > 25^\circ$ or $\sigma_\lambda > 25^\circ$ are shown at low opacity. Both the scatter plots and the histogram are colored by the host star's effective temperature, as in Figures \ref{fig:gamma_v_lambda} and \ref{fig:incl_v_lambda}.}
   \label{fig:inc_lambda_vs_sep}
\end{figure}

\subsection{Characterization of the clustering toward alignment}
\label{subsection:significance_alignment}

Next, we examined the robustness of the tentative \textit{over}abundance of fully aligned systems to quantify the likelihood that this population's alignment arises due to a physical mechanism, rather than by chance. More specifically, we hypothesize that the emergence of a population of fully aligned systems results from viscous dissipation at the systems' earlier protoplanetary disk phase \citep{bate2000observational, lubow2000tilting, foucart2014evolution, Zanazzi2018}. For simplicity, we consider only binary star systems within this section. All systems in the ``strict'' sample were included within this analysis, including those with large uncertainties $\sigma_i>25\degree$ and/or $\sigma_\lambda>25\degree$. Our conclusions remain unchanged when these systems are excluded.

We began by building a control sample of binary systems with properties similar to our sample, with the goal of quantifying the robustness of our results to biases that may arise from our orbit-fitting methods. For each system within our ``strict'' sample, we found the 10 most similar binary star systems in the \citet{el2021million} catalogue, adopting the similarity metric $M$ from \citet{christian2022possible}:

\begin{equation}
\begin{split}
    M = \Big(\frac{\Delta G_1}{4}\Big)^2 + \Big(\frac{\Delta G_2}{4}\Big)^2 + \Big(\frac{\Delta R_{p,1}}{4}\Big)^2 + \Big(\frac{\Delta R_{p,2}}{4}\Big)^2\\ + \Big(\frac{\Delta B_{p,1}}{4}\Big)^2 + \Big(\frac{\Delta B_{p,2}}{4}\Big)^2 + (\Delta\varpi)^2 + (\Delta s)^2.
\end{split}
\end{equation}
Here, $G$, $B_p$, and $R_p$ are the stellar magnitudes provided by \textit{Gaia} DR3, with subscripts for each of the two stars in the system. The projected separation and parallax of the binary system are given as $s$ and $\varpi$, respectively.

For this control sample, we implemented the same isochrone and orbit fitting process used in Section \ref{subsection:orbit_fitting} to derive orbital inclinations. For fifteen of the control sample systems, no solution was found with \texttt{lofti\_gaia} -- likely indicating that the systems are either unbound or that one source is a binary -- and the systems were discarded. These discarded systems were spread across our comparison sample, with no more than three failed cases from the control sample of any individual star. Therefore, we do not anticipate that removing these samples should significantly bias our results. 

\begin{figure}
    \centering
    \includegraphics[width=0.48\textwidth]{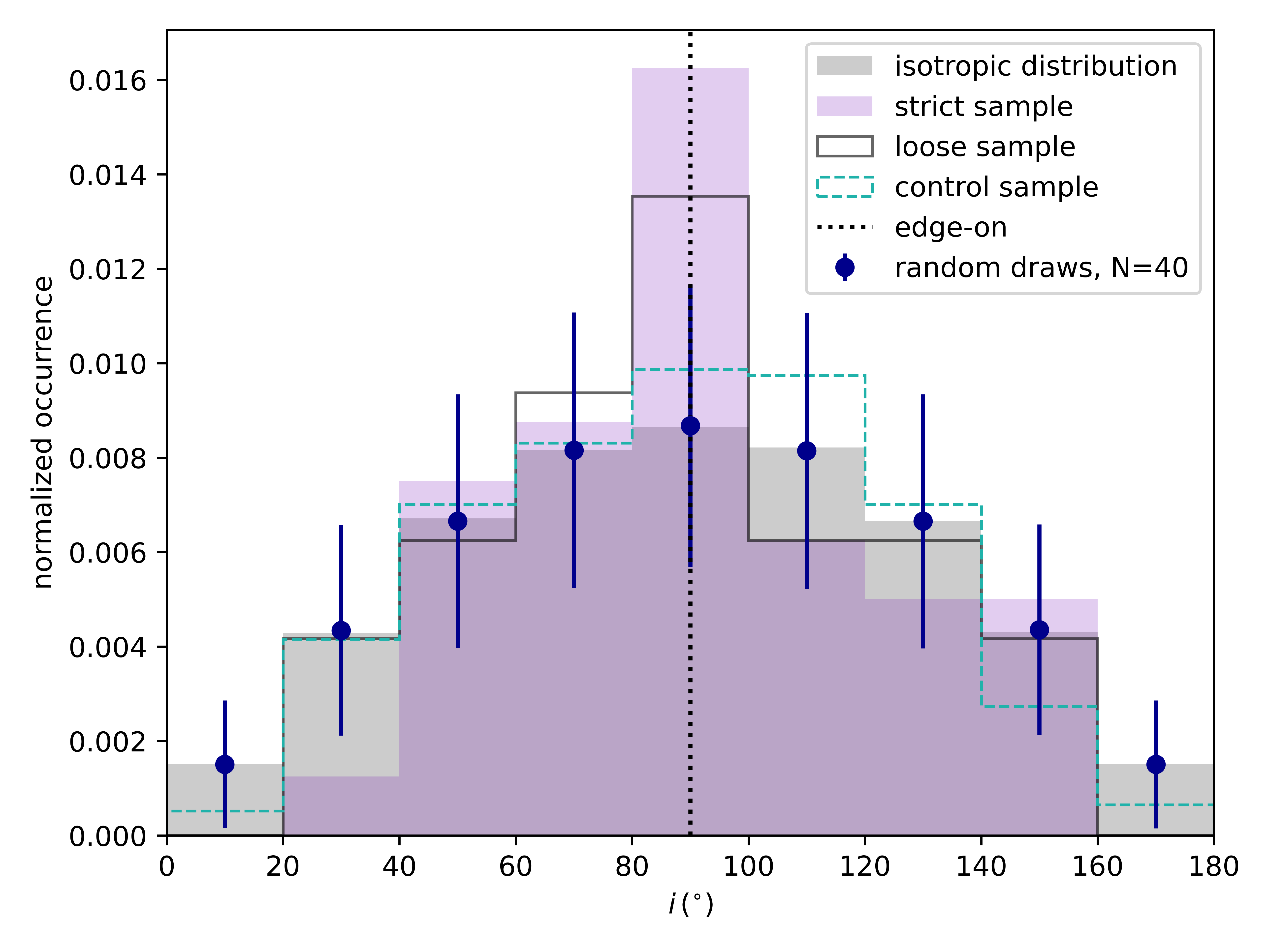}
    \caption{Distribution of observed stellar binary inclinations integrated over all stellar types, compared with inclinations drawn from an isotropic distribution. The ``strict'' sample is shown in purple, while the ``loose'' sample and the control sample are shown as black and green unfilled histograms, respectively. The mean and standard deviation of 150,000 random draws from an isotropic distribution, each containing 40 systems (the same number as our ``strict'' sample), are shown as dark blue error bars. }
    \label{fig:i_histograms_strict}
\end{figure}

The final control sample, which includes 385 systems in total, is shown in teal in Figure \ref{fig:i_histograms_strict} alongside the observed distribution of stellar binary inclinations -- showing the median values from our orbit fits -- and a reference isotropic distribution. In inclination space, $i=0\degree$ and $i=180\degree$ correspond to perfectly face-on orbits, while $i=90\degree$ corresponds to edge-on orbits. Although the control sample distribution lies entirely within the uncertainties derived from our random draws, its profile suggests some modeling bias toward near-edge-on inclinations relative to isotropy. This is an expected outcome of orbit fitting to astrometric observations with short orbital coverage -- particularly for orbits that may have significant, nonzero eccentricities -- as demonstrated in Section 4 of \citet{ferrer2021biases}.

To quantify the uncertainty associated with the relatively small size of the ``strict'' sample ($N=40$), we also derived the mean and standard deviation of 150,000 random draws from an isotropic distribution, where each draw contained 40 systems. The results of these random draws are provided as dark blue error bars in Figure \ref{fig:i_histograms_strict}. Based on the random draw distribution, the peak toward alignment observed across the full sample is a $2.5\sigma$ deviation from the uncertainties associated with the small sample size.

However, our primary observable of interest is not encapsulated solely by the peak toward edge-on orbits: instead, we aim to characterize the overall “peakiness” of the inclination distribution. Viscous dissipation should not only produce a peak toward alignment, but it should also push the full distribution toward a narrower set of inclinations about alignment. This is because, while some systems have already reached orbit-orbit alignment, others, in this framework, should have had their relative orbital inclinations damped (that is, pushed in the direction of orbit-orbit alignment) without reaching full alignment.



\begin{figure*}
    \centering
    \includegraphics[width = 0.7\linewidth]{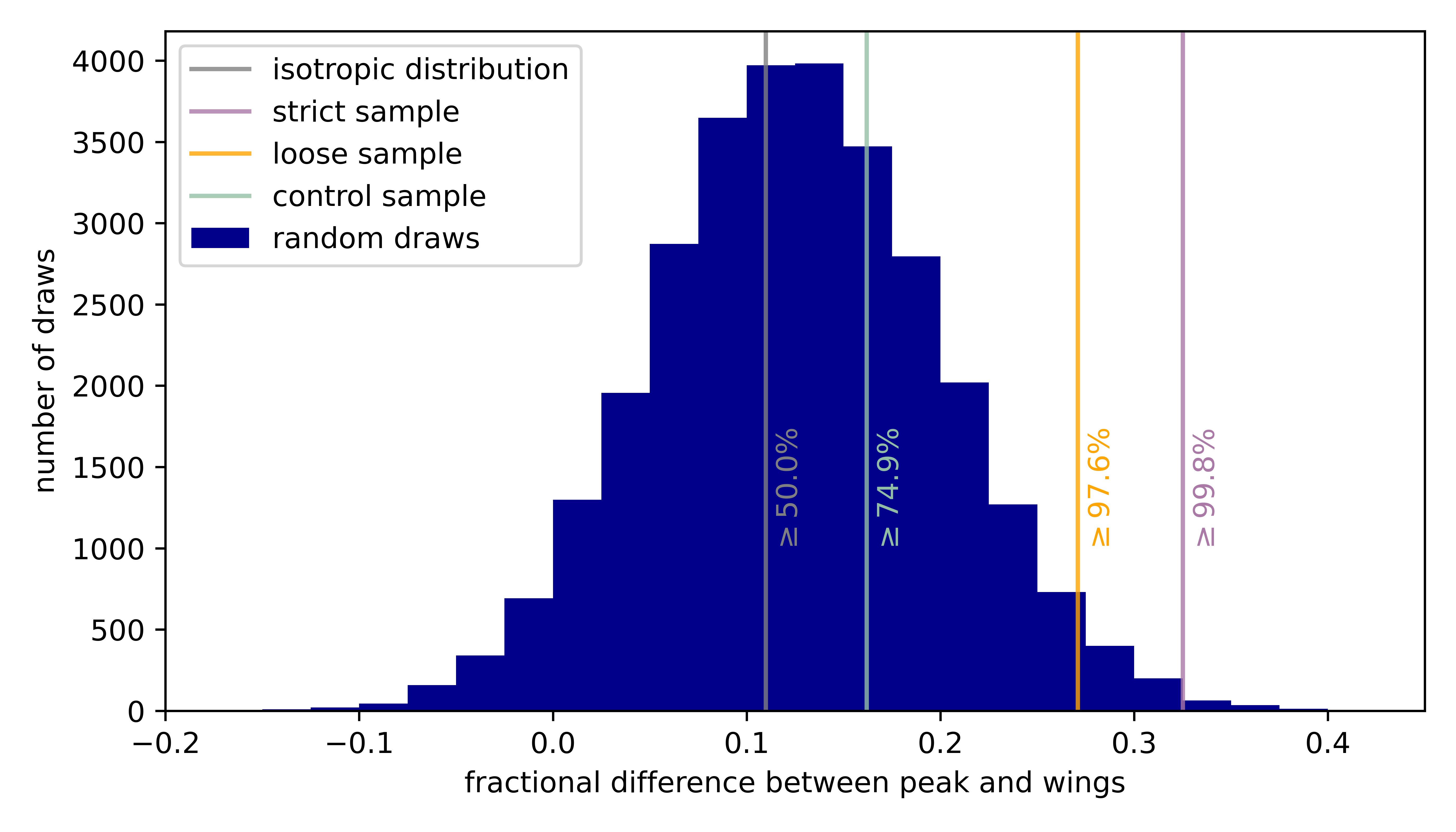}
    \caption{Distribution of the fractional difference $f$ between systems located within the $i=80-100\degree$ bin and those located in the $i=0-20\degree$ or $i=160-180\degree$ bins in Figure \ref{fig:i_histograms_strict}. The dark blue histogram is comprised of the same 30,000 random draws used to produce error bars in Figure \ref{fig:i_histograms_strict}. Based on the distribution of $f$, the ``strict'' sample is more centrally peaked toward edge-on than an isotropic distribution at the $\sim3\sigma$ level.} \label{fig:peak_diff_comparison}
\end{figure*}

Interestingly, in addition to the observed peak toward alignment, we also observe a tentative dearth of near-polar systems, manifesting as an absence of systems in the $i=0-20\degree$ and $i=160-180\degree$ bins in Figure \ref{fig:i_histograms_strict}. To quantify the “peakiness” of the observed distribution, we calculate the fractional difference $f$ between the number of systems $N_{80-100}$ that fall within the central bin of the distribution ($i=80-100\degree$) and the number of systems $N_{0-20}$ and $N_{160-180}$ that fall into the two bins that correspond to polar configurations, at $i=0-20\degree$ and $i=160-180\degree$, normalized by the total number of systems $N_{\mathrm{total}}$ across the full distribution:

\begin{equation}
    f = \frac{N_{80-100} - (N_{0-20} + N_{160-180})}{N_{\mathrm{total}}}.
\end{equation}
This fraction was calculated for each of our samples, as well as for each of the 30,000 random draws used to derive uncertainties in Figure \ref{fig:i_histograms_strict}. We intentionally exclude the bins between these extremes to avoid systems that are more likely to  undergo von Zeipel-Lidov-Kozai (ZLK) \citep{von_zeipel_1910, lidov1962evolution, kozai1962secular} oscillations (see Section \ref{subsection:inclinations_hot_cool} for further discussion on this topic). The results are shown in Figure \ref{fig:peak_diff_comparison}. 

The ``strict'' sample is more centrally peaked than an isotropic distribution at the $3\sigma$ level. The control sample also shows, at about the $1\sigma$ level, a low-level bias toward a peakier distribution than would be expected from isotropy. While this finding suggests that a portion of the ``peakiness'' observed within our sample may be attributable to modeling biases, the observed feature is a much stronger outlier than the deviation of the control sample.

\subsection{A broader spread of mutual inclinations for hot host stars}
\label{subsection:inclinations_hot_cool}

Another feature present in our sample is the apparent broader spread of orbit-orbit angles in systems with hot stellar hosts than in systems with cool stellar hosts, as shown in the top histogram of Figure \ref{fig:incl_v_lambda}. In the case of isotropy, systems with both cool and hot host stars should follow a distribution in binary inclination that is uniform in $\cos i$. By contrast, we observe a strong peak toward alignment in cool star systems ($\mu_{i, \mathrm{cool}}=100\degree$; $\sigma_{i, \mathrm{cool}}=21\degree$), with no clear peak and broader scatter for hot star systems ($\mu_{i, \mathrm{hot}}=86\degree$; $\sigma_{i, \mathrm{hot}}=37\degree$). This observation persists when stars with $\sigma_i>25\degree$ and $\sigma_{\lambda}>25\degree$ are excluded from the sample; in this case, $\mu_{i, \mathrm{cool}}=100\degree$; $\sigma_{i, \mathrm{cool}}=22\degree$; $\mu_{i, \mathrm{hot}}=89\degree$; and $\sigma_{i, \mathrm{hot}}=41\degree$.

The mutual inclination distribution for hot star systems does not appear random, but instead shows two suggestive peaks \textit{away} from isotropy, around $i=40-60\degree$ and $i=140-160\degree$. This double-peaked structure is reminiscent of the final mutual inclination distribution anticipated from the ZLK mechanism -- as seen in, for example, the bottom left panel of Figure 3 in \citet{naoz2012formation}. Therefore, we interpret this difference as evidence that, while cool star systems within our sample preferentially display orbit-orbit alignment, the hot star systems may have preferentially experienced ZLK oscillations.

\subsection{No clear correlation between spin-orbit and orbit-orbit misalignment}

We find no convincing correlation between spin-orbit and orbit-orbit alignment from our observed population. In particular, with a larger sample and updated parameters from \textit{Gaia} DR3, we do not recover the tentative correlation between spin-orbit and orbit-orbit misalignment suggested by \citet{behmard2022stellar}. Among the current sample of systems within both spin-orbit measurements and robust stellar companions, we find that systems that are spin-orbit misaligned span a wide range of orbit-orbit configurations, and vice versa.


\subsection{Significance of the observed trends in the viscous dissipation framework}
\label{subsection:nocorr}

Throughout all preceding sections, we considered the distribution of mutual orbital inclinations for all stellar companion separations. This choice was deliberate, to remain agnostic to the underlying mechanism producing the observed features in the distribution. Given the observed evidence in favor of (1) several fully aligned systems (Section \ref{subsection:aligned_systems}), (2) a relatively peaked distribution of binary inclinations (Section \ref{subsection:significance_alignment}), and (3) a wider spread in the inclination distribution for hot stars (Section \ref{subsection:inclinations_hot_cool}), we hypothesize that the inclination distribution for cool systems in our sample is shaped by viscous dissipation in nodally recessing protoplanetary disks that are warped by a nearby stellar perturber \citep{bate2000observational, lubow2000tilting, foucart2014evolution, Zanazzi2018}. We again consider only binary systems within this section for consistency with Section \ref{subsection:significance_alignment}, and we report results for the ``strict'' sample.

In the viscous dissipation framework, the set of close- to moderate-separation binaries would experience some inclination damping, while extremely wide-separation binaries would remain relatively unaffected and would instead serve primarily to add noise to our sample. To consider this scenario, we exclude extremely wide-separation binaries in this section, conservatively setting the outer limit for the influence of viscous dissipation to $s=2,000$ au. 

\begin{figure*}
    \centering
    \includegraphics[width = 0.7\linewidth]{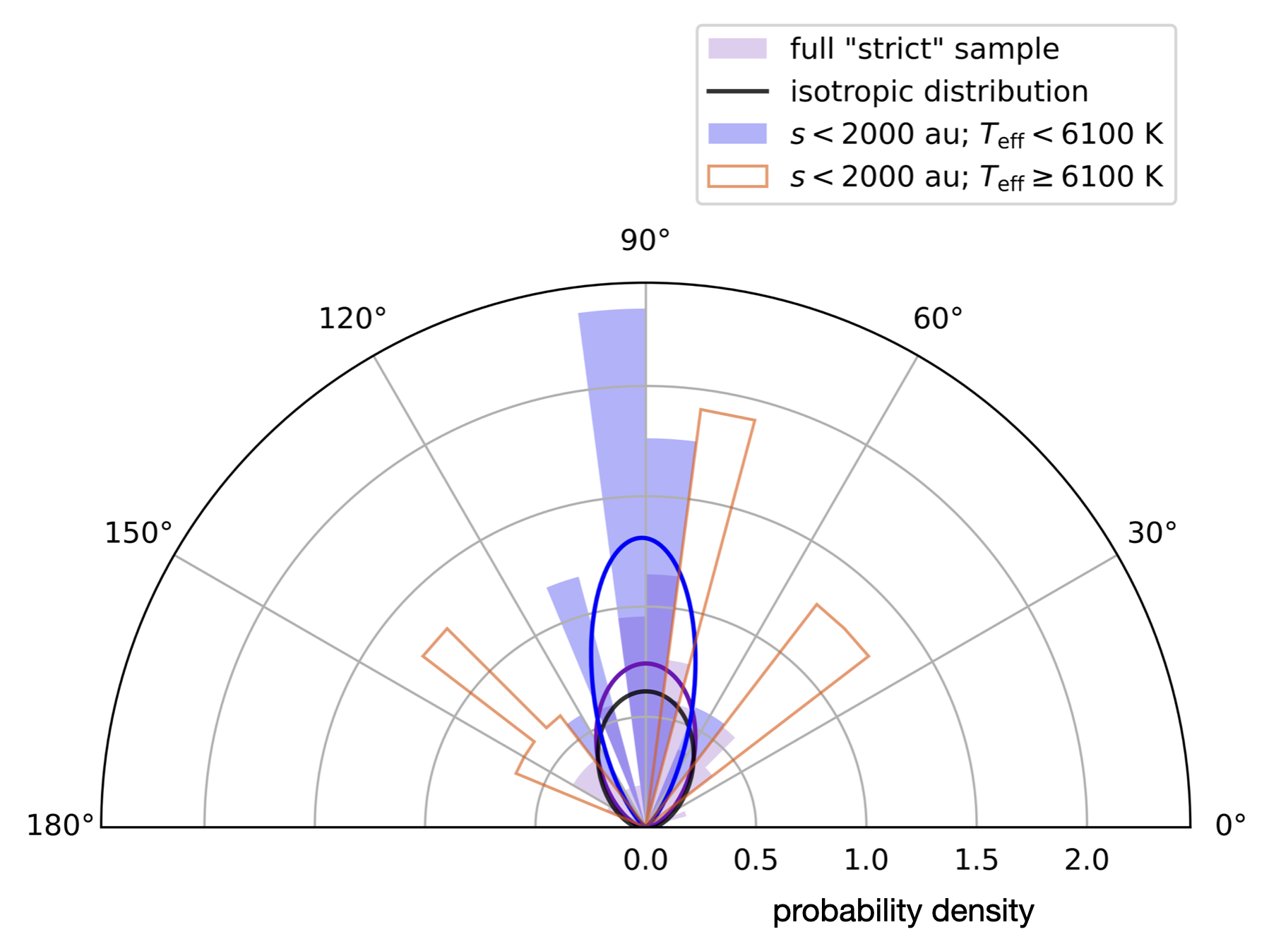}
    \caption{Normalized distribution of inclinations for the ``strict'' sample (purple filled histogram) with no $s$ or $T_{\mathrm{eff}}$ cuts, as well as the more restricted hot star (orange outline) and cool star (blue filled histogram) samples for comparison. Overplotted in the same colors are the maximum-likelihood von Mises probability density functions fitted to the full ``strict'' sample ($\mu=90.0\degree$; $\kappa=3.8$) and to the cool star subset sample ($\mu=90.8\degree$; $\kappa=11.1$), each of which demonstrates a single peak. The von Mises probability density function for a simulated isotropic distribution, with 30,000 draws as in Figure \ref{fig:i_histograms_strict}, is shown in black for reference, with $\mu=90.0\degree$ and $\kappa=2.7$.}\label{fig:polar_histogram_vonmises}
\end{figure*}

We applied the Kuiper statistic version of the Kolmogorov-Smirnov test to examine the significance of our observed trends for binaries with $s<2,000$ au. Specifically, we conducted one-sample Kuiper tests comparing the observed and/or control sample of binary inclinations to an isotropic distribution (uniform in $\cos i$), as well as two-sample Kuiper tests to compare subsets of the observed and/or control samples with each other. In each case, the null hypothesis -- that both samples are drawn from the same underlying distribution -- can be rejected if the $p$-value associated with the Kuiper test is $p < 0.05$. Each inclination distribution was characterized as $|\cos i|$ such that values fall within the range $0\leq|\cos i|\leq1$ and alignment corresponds to $|\cos i| = 0$, while polar orbits correspond to $|\cos i| = 1$. 

First, the observed sample and the control sample were compared to show whether they demonstrably arise from separate distributions. We drew from the control sample in 30,000 iterations to match the number of systems in each observed sample, then found the median of the $p$-values from all iterations. The two-sample Kuiper test accepts the null hypothesis that both samples are drawn from the same underlying distribution, with $p = 0.50$. This suggests that the underlying distributions do not strongly deviate from each other. The one-sample Kuiper test also accepts the null hypotheses that the full observed sample ($p = 0.65$) and the control sample ($p=0.28$) are each consistent with isotropy.


A series of Kuiper tests further affirms the observations outlined in Section \ref{subsection:inclinations_hot_cool}. A two-sample Kuiper test strongly rejects the null hypothesis that the cool and hot star systems within our sample stem from the same inclination distribution, with $p = 0.004$. When the same test is run between the hot and cool comparison stars in the control sample -- which includes binaries at a wide range of separations -- the null hypothesis is accepted with $p=0.39$. Furthermore, the one-sample Kuiper test rejects the null hypothesis that cool stars are drawn from an isotropic distribution ($p=0.04$), while it marginally accepts the null hypothesis that hot stars are drawn from an isotropic distribution ($p=0.09$). Therefore, although the full sample is consistent with isotropy, the two subsamples drawn from it appear to be distinct and marginally inconsistent with isotropy. Ultimately, we conclude that the cool and hot star systems in our sample demonstrate separate inclination distributions, with cool star systems showing a ``peakier'' distribution toward alignment, while hot star systems demonstrate broader scatter in their orbit-orbit orientations. 

The distributions of cool and hot star samples examined within this section are shown in Figure \ref{fig:polar_histogram_vonmises}, alongside the full ``strict'' sample for reference. While the hot star population, in orange, visibly deviates from a single overdensity, both the full sample and the cool star sample show peaks toward $i=90\degree$.  We fit the von Mises distribution, a wrapped Gaussian on a circle, to each of these two peaked distributions, following the probability density function

\begin{equation}
    f(x | \mu, \kappa) = \frac{\mathrm{exp}{(\kappa \cos(x-\mu))}}{2\pi I_0( \kappa)}
\end{equation}
where $\mu$ signifies the directionality of the distribution's peak; $\kappa$ is a measure of uniformity in which large values indicate higher concentration; and $I_0( \kappa)$ is the modified Bessel function of order zero. We find that the full ``strict'' sample is best fit by a von Mises probability density function with $\mu=90.0\degree$ and $\kappa=3.9$. When only the cool stars with nearby companions are considered, the location of the peak stays roughly static at $\mu=90.8\degree$, while the concentration toward the peak is much more pronounced ($\kappa=11.1$). For comparison, a von Mises fit to an isotropic distribution of inclinations is also shown, with a resulting peak at $\mu=90.0\degree$ with $\kappa=2.7$.

Finally, a Kuiper test confirms that, as suggested in Section \ref{subsection:nocorr}, no clear correlation is found between spin-orbit misalignment and orbit-orbit misalignment. A two-sample Kuiper test comparing the inclinations of systems with $\lambda > 25^\circ$ to systems with $\lambda \leq 25^\circ$ is accepted with $p = 0.58$.

\section{Discussion}
\label{section:discussion}

The results presented within this work suggest that viscous dissipation may be relatively efficient for some subset of planetary systems. While only eight of the 40 systems within our ``strict'' sample exhibit evidence of full alignment, the peaked distribution of inclinations suggests that many systems are pushed in the direction of alignment at some point during the system lifetimes. Fully aligned systems can be reproduced through strong viscous dissipation during the protoplanetary disk stage, together with star-disk coupling and an absence of resonant crossings or strong post-disk planet-planet interactions. 

Within the viscous dissipation framework, the widest-separation fully aligned systems identified in this work may represent chance orbit-orbit alignments in the line-of-sight direction. The HAT-P-1 and V1298 Tau systems have sky-projected separation $s=1,806$ au and $s=10,515$ au, respectively, between their two aligned components. Because the viscous damping rate scales as $\gamma_b\propto a_b^{-6}$ for true binary separations $a_b$ \citep{Zanazzi2018}, HAT-P-1 and V1298 Tau each have relatively long orbit-orbit alignment timescales by comparison with the other fully aligned systems in our sample. Given the prevalence of spin-orbit aligned systems within our sample, a joint spin-orbit and line-of-sight orbit-orbit alignment for one or both systems could plausibly arise by chance from the independent evolution of the two stars in each of the systems.

The broader range of inclinations observed in hot star systems as compared with cool star systems may also offer dynamical insights into the underlying properties of protoplanetary disk systems. Within the viscous dissipation framework, the observed difference in the inclination distribution could arise due to systematic differences in disk extent, aspect ratio, and viscosity. For example, protoplanetary disks around low-mass stars may have a lower temperature and therefore lower aspect ratio, enabling them to dissipate energy more efficiently than disks around high-mass stars. Differences in the protoplanetary disk dispersal timescales for cool and hot star systems may produce systematic differences in their final geometries, as well: previous work has shown that high-mass stars have systematically shorter protoplanetary disk lifetimes than lower-mass stars \citep{ribas2015protoplanetary}, likely due to differences in the dispersal rate of disk material through viscous accretion, magnetohydrodynamic winds, and photoevaporation as a function of stellar type \citep{komaki2021radiation, komaki2023simulations}. Hot star systems would, therefore, have less time available to realign prior to protoplanetary disk dispersal.

We emphasize that, because of the relatively deep transits that are necessary to resolve the Rossiter-McLaughlin effect or Doppler tomography signals, our sample is dominated by systems that host short-period giant planets. As a result, our sample is not representative of the full breadth of system types. Instead, the presented results are biased toward systems that likely had relatively high-mass and/or high-metallicity natal protoplanetary disks. Further joint spin-orbit and orbit-orbit analyses for systems with smaller, lower-mass planets are necessary to characterize the generality of the trends examined within this work.

\section{Conclusions}
\label{section:conclusions} 

In this work, we examined the joint spin-orbit and orbit-orbit distribution for exoplanets residing in binary and triple-star systems. We considered the full sample of systems with both (1) an exoplanet spin-orbit measurement within the TEPcat catalogue and (2) one or more stellar companions resolved by \textit{Gaia} DR3. Our conclusions are as follows:

\begin{itemize}
    \item We identify a population of nine exoplanet host systems that each exhibit evidence of joint spin-orbit and orbit-orbit alignment. The joint alignment of seven of these systems was demonstrated for the first time: six binary star systems (CoRoT-2, DS Tuc, HAT-P-1, HAT-P-22, HD 189733, and TrES-4) were found to exhibit evidence of full system alignment, and one triple-star system (V1298 Tau) exhibits evidence of full alignment between the host star's transiting exoplanet companion and one of the two companion stars. 
    \item The sample of stellar binary inclinations for exoplanet-hosting systems with spin-orbit constraints is more strongly peaked toward alignment than an isotropic distribution -- a trend that appears to be driven largely by cool host star ($T_{\mathrm{eff}}<6100$ K) systems within the sample. We find a joint overabundance of systems demonstrating line-of-sight orbit-orbit alignment and underabundance of face-on systems that correspond to near-polar binary orbital configurations. This observed trend may be a signature of viscous dissipation induced by nodal recession during the protoplanetary disk phase.
    \item For the set of systems with $s<2,000$ au, we demonstrate a broader spread in orbit-orbit orientations observed in hot star systems than in cool star systems. Eight of the nine observed fully aligned systems also include a cool host star below the Kraft break. In the viscous dissipation framework, these observations provide evidence that protoplanetary disks around cool stars may be longer-lived and/or more efficient at dissipating energy than those around hot stars.
    \item No clear correlation is found between spin-orbit misalignment and orbit-orbit misalignment. This result indicates that, while mechanisms related to spin-orbit excitation may, in some cases, be enhanced by orbit-orbit misalignments, orbit-orbit misalignment does not necessarily accompany spin-orbit misalignment (and vice versa).
\end{itemize}

Our results highlight an observed regime within which moderately wide-separation stellar companions push their neighbors' protoplanetary disks, and therefore the planetary systems that emerge from those disks, toward an aligned state, approaching a more general three-body equilibrium \citep{hut1981tidal} and enhancing order in exoplanet systems. Further observations of similar, well-constrained systems offer the potential to provide precise dynamical limits on protoplanetary disk properties, offering a new lens through which disk properties -- including masses, viscosities, and lifetimes -- may be constrained.

\section{Acknowledgements}
\label{section:acknowledgements}

We thank J. J. Zanazzi, Songhu Wang, Sam Christian, Fei Dai, Aida Behmard, Smadar Naoz, and Matthew Bate for helpful discussions over the course of this work. M.R. acknowledges support from Heising-Simons Foundation Grant \#2023-4478, as well as the 51 Pegasi b Fellowship Program. This research has made use of the NASA Exoplanet Archive, which is operated by the California Institute of Technology, under contract with the National Aeronautics and Space Administration under the Exoplanet Exploration Program. This work has made use of data from the European Space Agency (ESA) mission {\it Gaia} (\url{https://www.cosmos.esa.int/gaia}), processed by the {\it Gaia} Data Processing and Analysis Consortium (DPAC, \url{https://www.cosmos.esa.int/web/gaia/dpac/consortium}). Funding for the DPAC has been provided by national institutions, in particular the institutions participating in the {\it Gaia} Multilateral Agreement.

\software{\texttt{numpy} \citep{oliphant2006guide, walt2011numpy, harris2020array}, \texttt{matplotlib} \citep{hunter2007matplotlib}, \texttt{pandas} \citep{mckinney2010data}, \texttt{lofti\_gaia} \citep{pearce2020orbital}, \texttt{isoclassify} \citep{huber2017isoclassify, huber2017asteroseismology, berger2020gaia}, \texttt{mwdust} \citep{bovy2016galactic}, \texttt{scipy} \citep{virtanen2020scipy}, \texttt{astropy} \citep{astropy2013, astropy2018, astropy2022}}

\facility{NASA Exoplanet Archive, Extrasolar Planet Encyclopaedia, TEPcat}

\appendix

\section{Joint Alignment Candidates}
\label{section:joint_alignment_candidates}

While building the final samples for this study, we identified several candidates for joint spin-orbit and orbit-orbit alignment that nearly satisfy the four robustness tests described in Section \ref{subsection:aligned_systems}, but that were removed from the final sample in this work due to less well-constrained orbits that manifest as relatively large uncertainties. We classify these systems as ``joint alignment candidates'' and discuss their properties here.

Several joint alignment candidates were identified from the sample of binary star systems examined within this work. WASP-26 satisfies Conditions 1, 2, and 3 but has too weak a spin-orbit constraint ($\lambda=-34^{+36}_{-26}\degree$) to conclusively determine whether the host star's companion exoplanet is spin-orbit aligned. WASP-87 meets all conditions except Condition 2, such that the system may be near, but not exactly, edge-on in $\gamma$. Four further systems (HAT-P-24, TrES-1, WASP-18, and WASP-78) meet Conditions 1 and 3 but have relatively large uncertainties such that they are not unambiguously edge-on in either $\gamma$ or $i$. Further astrometric monitoring or radial velocity measurements between the two binary components would help to conclusively determine whether these four systems are edge-on.


One additional triple-star system, WASP-24, hosts an aligned planetary companion and may be consistent with orbit-orbit alignment. Only two of the three stars in the WASP-24 system are resolved in \textit{Gaia} DR3: the secondary component was identified as an eclipsing binary in \citet{street2010wasp}. However, the large uncertainty in $\gamma$ between the two resolved stars within the WASP-24 system ($\sigma_{\gamma}=55.5\degree$) renders the measurement of $\gamma$ relatively uninformative, such that the system's true alignment remains ambiguous.

All joint alignment candidates discussed in this section are present in both the ``strict'' and ``loose'' samples. No additional joint alignment candidates were found in the ``loose'' sample that are not present in the ``strict'' sample.

\section{Triple-Star Systems}
\label{section:triple_systems}
Five triple-star systems emerged from our sample vetting: K2-290, Kepler-13, V1298 Tau, WASP-11, and WASP-24. In this section, we discuss the geometric properties of these systems in greater detail.

Of the five systems identified as triple-star systems within this work, four had both stellar companions identified in previous publications. WASP-11 and WASP-24 were shown to be bound
 triple-star systems by \citet{mugrauer2019search}, while Kepler-13 was identified as a triple-star system by \citet{santerne2012sophie} and K2-290 was identified as a triple-star system in the discovery paper for its two companion planets \citep{hjorth2019k2}. Each of these four systems has only two stellar components resolved by \textit{Gaia} DR3, likely due to the relatively close separation of the third component in each case. Kepler-13 BC is a spectroscopic binary \citep{johnson2014misaligned}, while WASP-24 BC is an eclipsing binary \citep{street2010wasp, mugrauer2019search}. K2-290 B is located at a separation of only $0.4\arcsec$ from the primary star \citep{hjorth2019k2} and HAT-P-10 B is only $0.36\arcsec$ from the primary star \citep{ngo2015friends}, such that both stars are likely unresolved by \textit{Gaia} DR3. 

V1298 Tau, by contrast, has not been previously confirmed as a triple-star system. Of the five identified systems, V1298 Tau is also the only one that has all three stellar components resolved by \textit{Gaia} DR3. \citet{david2019warm} conducted several robustness tests searching for close-in companions but identified none. Our results are consistent with the absence of nearby ($s<10,000$ au) stellar companions in the V1298 Tau system.

We observe that the more nearby stellar companion to V1298 Tau, at $s=10,515$ au, is consistent with alignment ($\gamma=4.2\pm3.2\degree$), while the further companion at $s=14,632$ au is likely misaligned at $\gamma=125.5\pm4.4\degree$. The close alignment of such a distant companion in this system is surprising and challenging to reproduce through dissipative channels over the expected lifetime of a protoplanetary disk. Therefore, it is possible that the line-of-sight orbit-orbit alignment observed in the V1298 Tau system may reflect a chance alignment.


All resolved companions in the identified triple-star systems were identified at relatively wide sky-projected separations $s>2,000$ au from the primary, with the exception of Kepler-13, which has a secondary companion at $s=552$ au. Kepler-13 b is spin-orbit misaligned ($|\lambda|=58.6\pm2.0\degree$) and the secondary star has $\gamma=147.2\pm7.6\degree$ relative to the primary, indicating that the secondary star's orbital plane is likely not edge-on.


\bibliography{bibliography}
\bibliographystyle{aasjournal}

\end{document}